\documentstyle[12pt,epsfig,axodraw]{article}

\newcommand{\s}{\smallskip}

\newcommand{\lsim}{\raisebox{-0.13cm}{~\shortstack{$<$ \\[-0.07cm] $\sim$}}~}
\newcommand{\gsim}{\raisebox{-0.13cm}{~\shortstack{$>$ \\[-0.07cm] $\sim$}}~}

\newcommand{\nn}{\noindent}
\newcommand{\non}{\nonumber}
\newcommand{\fll}{\begin{flushleft}}
\newcommand{\flr}{\begin{flushright}}
\newcommand{\beq}{\begin{eqnarray}}
\newcommand{\eeq}{\end{eqnarray}}

\newcommand{\ctw}{c_W}
\newcommand{\stw}{s_W}
\newcommand{\ctwn}[1]{c^#1_W}
\newcommand{\stwn}[1]{s^#1_W}

\newcommand{\ee}{e^+e^-}
\newcommand{\tb}{\tan \beta}
\newcommand{\p}[1]{p_{_#1}}
\newcommand{\pn}[1]{\not{p}_#1}
\newcommand{\charpmi}[0]{\chi^\pm_i}
\newcommand{\charpmj}[0]{\chi^\pm_j}
\newcommand{\azff}{a_{Z \tilde{f} \tilde{f}}}

\newcommand{\dotp}[2]{p_{_{#1#2}}}
\newcommand{\qf}{Q_f}
\newcommand{\qx}{Q_\chi}
\renewcommand{\ae}{a_e}
\newcommand{\ve}{v_e}
\newcommand{\af}{a_f}
\newcommand{\vf}{v_f}
\newcommand{\aih}{A_{ih}}
\newcommand{\ajh}{A_{jh}}
\newcommand{\akh}{A_{kh}}

\newcommand{\olj}{O_{ij}^{h^\prime}}
\newcommand{\olk}{O_{ik}^{h^\prime}}
\newcommand{\orj}{O_{ij}^h}
\newcommand{\ork}{O_{ik}^h}

\newcommand{\msfh}{m_{{\tilde{f}_h}}}
\newcommand{\mfer}{m_{f}}
\newcommand{\mz}{m_{Z}}
\newcommand{\mgi}{m_{\chi_i}}
\newcommand{\mgj}{m_{\chi_j}}
\newcommand{\mgk}{m_{\chi_k}}

\begin{document}

\vspace*{.1cm} 
\baselineskip=17pt

\begin{flushright}
PM/01--37\\
November 2001\\
\end{flushright}

\vspace*{0.9cm}

\begin{center}

{\large\sc {\bf Associated production of sfermions and gauginos}}

\vspace*{0.4cm}

{\large\sc {\bf at high--energy e$^+$e$^-$ colliders}}

\vspace{0.7cm}

{\sc Aseshkrishna DATTA} and {\sc  Abdelhak DJOUADI} 

\vspace{0.7cm}

Laboratoire de Physique Math\'ematique et Th\'eorique, UMR5825--CNRS,\\
Universit\'e de Montpellier II, F--34095 Montpellier Cedex 5, France. 

\end{center} 

\vspace*{1cm} 

\begin{abstract}

\nn In the context of the Minimal Supersymmetric extension of the Standard
Model, we analyze the production at future high--energy $\ee$ colliders of
second and third generation scalar leptons as well as scalar quarks in
association with neutralinos and charginos, $\ee \to f \tilde{f} \chi$. In the
case of third generation squarks, we also discuss the associated production
with gluinos.  We show that the cross sections for some of these three--body
final state processes could be significant enough to allow for the detection of
scalar fermions with masses above the kinematical two--body threshold,
$\sqrt{s}= 2m_{\tilde{f}}$.  We then discuss, taking as a reference example the
case of scalar muons, the production cross sections in various approximations
and make a comparison with the full four--body production process, $\ee \to f
\bar{f} \chi \chi$, in particular around the two--sfermion threshold.  

\end{abstract}

\newpage 

\subsection*{1. Introduction}

The search for Supersymmetric (SUSY) particles is one of the major goals of
future high--energy colliders. Because of their strong interactions, scalar
quarks and gluinos can be best searched for at hadron machines such as the
Tevatron \cite{Tevatron} and the LHC \cite{LHC}, where they are copiously
produced. On the other hand, the weakly interacting particles, the left-- and
right--handed scalar leptons, $\tilde{\ell}_L$ and $\tilde{\ell}_R$, as well as
the charginos and the neutralinos, $\chi_{1,2}^\pm$ and $\chi_{1-4}^0$, can be
more efficiently probed at high--energy $\ee$ colliders where the signals are
cleaner and the backgrounds are smaller \cite{NLC,TESLA}. The cleaner
environment also allows for the detailed study of the properties of these
particles and provides the possibility to reconstruct parts of the SUSY
Lagrangian at the low energy scale, opening a window for the determination of
the structure of the theory at the very high energy scale \cite{tests}. \s

In the Minimal Supersymmetric extension of the Standard Model (MSSM)
\cite{MSSM,HaberKane}, the lightest SUSY particle (LSP) is the lightest
neutralino $\chi_1^0$. If R--parity is conserved, this particle is stable and
since it is electrically neutral, it is invisible and escapes experimental
detection [at least in the simplest pair production processes].  In models
where the gaugino masses are unified at the GUT scale, as in the minimal
Supergravity model (mSUGRA)  \cite{mSUGRA}, the masses of the lightest
charginos and neutralinos are such that: $m_{\chi_2^0} \sim m_{\chi_1^\pm} \sim
2 m_{\chi_1^0}$ in the case where they are gaugino--like or $m_{\chi_2^0} \sim
m_{\chi_1^\pm} \sim m_{\chi_1^0}$ in the case where they are higgsino--like. 
Thus, the states $\chi_2^0$ and $\chi_1^+$ are not much heavier than the LSP
and might be the first SUSY particles to be discovered.  In models with a
common scalar mass at the GUT scale, sleptons (in particular $\tilde{\ell}_R$)
are in general much lighter than squarks and can have masses comparable to
those of the lighter neutralinos and charginos. In the case of the
$\tilde{\tau}$ slepton, mixing effects that are proportional to the mass of the
partner fermion, generate a possibly large splitting between the mass
eigenstates, making that one of them is much lighter than the other
sleptons. [The mixing in the case of selectrons and smuons, as well as
for first and second generation squarks, is extremely small and can be safely
neglected.]  The scalar quarks of the third generation, $\tilde{Q}= \tilde{t}$
and $\tilde{b}$, because of the large Yukawa couplings of their partners
quarks, can also mix strongly leading to one mass eigenstate which is rather
light.  These particles can therefore be among the first ones to be accessible
at the next generation of $\ee$ colliders. \s

In $\ee$ collisions, left--and right--handed scalar leptons can be 
produced in pairs \cite{Asleptons}, 
\beq
\ee \to \tilde{\ell}_i \tilde{\ell}_i^* \ \ \  (i=L,R) 
\eeq
if the center of mass energy is high enough, $\sqrt{s} > 2 m_{\tilde{\ell}}$. 
In the case of second and third generation sleptons, the production mechanisms
proceed through $s$--channel gauge boson exchange, $\gamma$ and $Z$--boson for
charged sleptons and only $Z$--boson for the sneutrinos. Mixed pairs of left--
and right--handed charged sleptons [there is no right--handed sneutrino in the
MSSM] cannot be produced in $\ee$ collisions, since $\gamma$ and $Z$ bosons do
not couple to $\tilde{\ell}_L \tilde{\ell}_R^*$ states. For first generation
sleptons, additional channels are provided by the $t$--channel exchange of
neutralinos in the case of selectrons and charginos in the case of sneutrinos,
making the situation slightly more complicated. For third generation squarks,
the pair production of the lightest states, $\ee \to \tilde{t}_1 \tilde{t}_1^*$
and $\tilde{b}_1 \tilde{b}_1^*$, proceeds only through $s$--channel $\gamma$ 
and $Z$--boson exchange \cite{Asquarks}. \s

Far above the kinematical threshold, the production of charged sfermions
[ignoring the $t$--channel exchange diagrams] is dominated by the photon
exchange and the cross sections, normalized to the QED point like cross section
for muon pair production $\sigma_0=4\pi \alpha^2/(3s)$, are simply given by
$R_{\tilde{f}} \sim N_{\tilde{f}} Q^2_{\tilde{f}} /4$, where $Q_{\tilde{f}}$ is
the electric charge and $N_{\tilde{f}}$ the color factor ($=3$ for squarks and
$1$ for sleptons) of the sfermion $\tilde{f}$, and the factor $1/4$ comes from
the absence of spin sum for scalar particles compared to the case of final
fermions. The cross sections are therefore rather large: in the case of smuons,
one has $\sigma(\ee \to \tilde{\mu}^+_i \tilde{\mu}^-_i) \sim 100$ fb at
a c.m.  energy $\sqrt{s}=500$ GeV, leading to a sample of the order of 50 000
events with the high integrated luminosities, $\int {\cal L} \sim 500$
fb$^{-1}$, expected at future linear colliders [as is the case for the TESLA
machine \cite{TESLA} for instance]. This large number of events will therefore
allow very detailed studies of the properties of these particles. \s

Due the P--wave nature of the production mechanisms for scalar particles [we
again ignore the $t$--channel contributions], the cross sections are strongly
suppressed by $\beta_{\tilde{f}}^3= (1- 4 m_{\tilde{f}}^2/s)^{3/2}$ factors
near the kinematical thresholds. Scanning near these thresholds is very
important since it allows us to determine accurately the masses of the produced
sfermions \cite{TESLA}. In this case, a very refined analysis of the cross 
sections \cite{Peter,Manuel}, including the non--zero decay width of the 
produced states and some higher order effects [such as Coulomb re--scattering 
effects and initial or final state radiation], has to be performed.  \s

An interesting question to ask is: what if the center of mass energy is slightly
below the kinematical threshold for sfermion pair production, $\sqrt{s} \lsim
2 m_{\tilde{f}}?$ In this case, one has to consider the production of 
off--shell sfermions, subsequently decaying into neutralinos or charginos and
fermions. In fact, in the general case, one has to consider the associated 
production of one sfermion, its partner fermion and a light neutralino or 
a chargino state [and only in the case where they are lighter than the 
sfermion as is the case of the neutralino LSP]. These are three--body 
production processes and the cross sections should be suppressed by an extra 
power of the electroweak coupling $\alpha_{\rm EW}$ and by the virtuality of 
the exchanged (s)particles. However, since the pair production cross sections 
are rather large, one might hope that the three--body cross sections are not 
too small if one is not far above the two--body kinematical production 
threshold. \s

\nn These higher order processes are worth studying for at least four good
reasons: 
\begin{itemize} 
\item[$(i)$] They provide the possibility to discover sfermions even if they
are too heavy to be produced in pairs, i.e.  they increase the discovery reach
of the $\ee$ colliders.  
\item[$(ii)$] In the case where the sfermion is the next--to--lightest
sparticle, it will decay 100\% of the time into its partner fermion and the LSP
and the information on the sfermion--fermion--neutralino coupling is lost;
reducing the energy below threshold where one has only the three--body process,
allows an access to this coupling.  
\item[$(iii)$] To make a threshold scan for the measurement of the masses one
has to include the non--zero decay widths of the sfermions, i.e. to consider
the four--body production process of two fermions and two gauginos; it is then
interesting to compare the cross section in this more complicated case with the
one of the three-body production process, i.e. with only one resonant sfermion,
which is easier to handle.  
\item[$(iv)$] In the strong interaction sector, it provides access to the
gluino which, if it is heavier than squarks, cannot be produced otherwise in
$\ee$ collisions [except via loops and the cross sections in this case are too
small].  
\end{itemize}

In this paper, we therefore analyze, in the context of the MSSM, the associated
production of sfermions with neutralinos and charginos (and in the case of
third generation squarks, with gluinos) at future high--energy $\ee$ colliders.
For sleptons, we consider the case of left-- and right--handed smuons, stau's
(for which the mixing will be included), as well as their partners sneutrinos;
the more complicated case of associated production of $\tilde{e}$ and
$\tilde{\nu}_e$ with gauginos will be discussed in a forthcoming paper
\cite{later}. For squarks, we will discuss the associated production of the
lightest states $\tilde{t}_1$ and $\tilde{b}_1$ with neutralinos, charginos as
well as with gluinos. The amplitudes for these processes will be given and the
cross sections for masses above the two--body kinematical threshold will be
shown. In the case of $\tilde{\mu}_R, \tilde{\mu}_L$ and $\tilde{\nu}_\mu$, we 
will also discuss in details some approximations and the behavior of the cross 
sections near the kinematical thresholds. \s

The remainder of the paper is organized as follows. In the next section, we
discuss the sfermion--gaugino associated production mechanisms and present the 
amplitudes for the various channels. In section 3, we give some numerical 
illustrations for the production of second and third generation 
neutral and charged sleptons as well as for third generation squarks, 
including the associated production with gluinos). In section 4, we discuss 
various approximations and the threshold behavior of the cross sections.
A short conclusion is given in section 5.  In the Appendix, we present the 
analytical expression of the differential cross section for the associated 
production in the very good approximation where only the $s$--channel  
contributions are taken into account.

\subsection*{2. Production mechanisms and amplitudes}

\subsubsection*{2.1 The Feynman diagrams for the various processes}

The Feynman diagrams contributing to the associated production in $\ee$
collisions of squarks and second/third generation sleptons with the lighter
chargino $\chi_1^\pm$ and neutralinos $\chi_{1,2}^0$, that we will call
electroweak gauginos\footnote{In the range of parameter space where the
production cross sections are favorable, the lighter charginos and neutralinos
will be indeed gaugino--like as will be discussed later. We will not consider
the case of the associated production with the heavier chargino and neutralinos
since the cross sections will be suppressed, at least by the smaller
phase--space. Furthermore, for third generation sleptons and for squarks, we 
will consider only the more favorable cases of the lighter $\tilde{\tau}_1$ 
as well as $\tilde{t}_1$ and $\tilde{b}_1$ states.} for short, are displayed 
in Fig.1. More explicitly, the various possibilities which will be studied in 
this paper are, for sleptons
\beq 
\ee & \to & \tilde{\mu}_i^\pm \mu^\mp \chi_{1,2}^0 \ , \ \tilde{\mu}^\pm_i 
\nu_{\mu} \chi_1^\mp \ \ \ {\rm with}~i=L,R \non \\
\ee & \to & \tilde{\nu}_\mu \mu^\pm \chi_1^\mp \ , \ \tilde{\nu}_\mu 
\nu_\mu \chi_{1,2}^0 \non \\
\ee & \to & \tilde{\tau}_i^\pm \tau^\mp \chi_{1,2}^0 \ , \ \tilde{\tau}^\pm_i 
\nu_{\tau} \chi_{1}^\mp \ \ \ \ {\rm with}~i=1 \non \\
\ee & \to & \tilde{\nu}_\tau \tau^\pm \chi_1^\mp \ , \ \tilde{\nu}_\tau 
\nu_\tau \chi_{1,2}^0 
\eeq
and for squarks [the diagrams for gluino final states are shown in Fig.~2]
\beq 
\ee & \to & \tilde{t}_1 \bar{t} \chi_{1,2}^0 \ , \ \tilde{t}_1 \bar{b} \chi_1^- 
\ , \ \tilde{t}_1 \bar{t} \tilde{g} \non \\
\ee & \to & \tilde{b}_1 \bar{b} \chi_{1,2}^0 \ , \ \tilde{b}_1 \bar{t} \chi_1^+ 
\ , \ \tilde{b}_1 \bar{b} \tilde{g}
\eeq
In the case of smuon final states, the production is mediated by $s$--channel
$\gamma$ and $Z$ boson exchanges with either a smuon (1a) or a muon (1b) is
virtual and goes into a (charged/neutral) lepton or slepton and a
(neutral/charg)ino.  Additional contributions come from the production of
chargino and neutralino pairs, either through $s$--channel gauge boson exchange
($\gamma,Z$ for $\chi_i^\pm$ and only $Z$ for $\chi_i^0$, Fig.~1c) or through
$t$--channel first generation slepton exchange: $\tilde{\nu}_e$ for
$\chi_i^\pm$, Fig.~1d or 1d' (depending on the isospin of the final sfermion)
and $\tilde{e}_{L,R}$ for $\chi_i^0$ final states, Fig.~1d and Fig.~1e (note
that here, there are both $t$-- and $u$--channels because of the Majorana
nature of the neutralinos) with the virtual gaugino going into a smuon/neutrino
or smuon/muon final states, respectively.  Note that in Fig.~1c, there is no
additional diagram where the slepton--lepton pair is emitted from the other
neutralino since the gauge boson-$\chi_i^0$-$\chi_j^0$ vertex is already 
symmetrized. \s

In the case of sneutrino final states, the same diagrams in Fig.~1 contribute
except for the fact that there is no $\gamma$ exchange in diagrams (1a) and
(1b).  For $\tilde{\tau}_1$ final states, an additional complication comes from
the mixing which has to be taken into account in the
$\chi_i^+$--$\tilde{\tau}$--$\nu_\tau$ $\chi_i^0$--$\tilde{\tau}$-$\tau$ and
$Z$--$\tilde{\tau}$--$\tilde{\tau}$ vertices and the exchange of the heavier
$\tilde{\tau}_2$ slepton in Fig.~1a when the $Z$ boson is exchanged.  \s

The case of associated production of third generation squarks with their 
partners quarks and electroweak gauginos is to be handled similarly as the case
of $\tilde{\tau}$ sleptons, i.e. to include the mixing in the various couplings 
and the exchange of the heavier squark states in diagram (1a) when 
the $Z$--boson is involved. For the associated production of squarks and 
gluinos, the situation is relatively simpler since only two diagrams are 
involved: squark and quark pair production with the virtual squark or quark
going into a pair of quark/gluino or squark/gluino; Fig.~2.  \bigskip

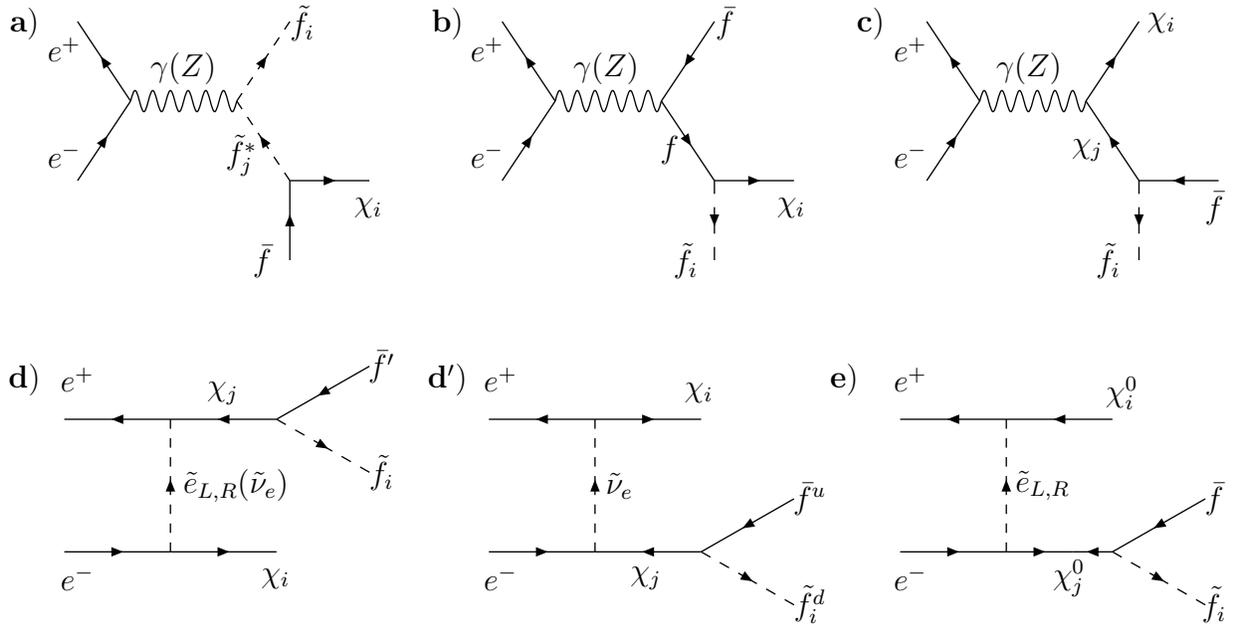
\begin{figure}[htbp]
\vspace*{.5cm}
\begin{picture}(1000,180)(0,0)
\Text(0,180)[]{${\bf a)}$}
\Text(15,130)[]{$e^-$}
\Text(15,170)[]{$e^+$}
\ArrowLine(20,120)(40,150)
\ArrowLine(40,150)(20,180)
\Photon(40,150)(80,150){4}{6}
\Text(60,163)[]{$\gamma (Z)$}
\DashArrowLine(100,120)(80,150){4}
\DashArrowLine(80,150)(100,180){4}
\Text(82,130)[]{$\tilde{f}_j^*$}
\Text(105,180)[]{$\tilde{f}_i$}
\ArrowLine(100,120)(130,120)
\ArrowLine(100,90)(100,120)
\Text(130,110)[]{$\chi_i$}
\Text(90,90)[]{$\bar{f}$}
\Text(160,180)[]{${\bf b)}$}
\Text(175,130)[]{$e^-$}
\Text(175,170)[]{$e^+$}
\ArrowLine(180,120)(200,150)
\ArrowLine(200,150)(180,180)
\Photon(200,150)(240,150){4}{6}
\Text(220,163)[]{$\gamma (Z)$}
\ArrowLine(260,180)(240,150)
\ArrowLine(240,150)(260,120)
\Text(244,132)[]{$f$}
\Text(265,180)[]{$\bar{f}$}
\DashArrowLine(260,120)(260,90){4}
\ArrowLine(260,120)(290,120)
\Text(290,110)[]{$\chi_i$}
\Text(250,90)[]{$\tilde{f}_i$}
\Text(320,180)[]{${\bf c)}$}
\Text(335,130)[]{$e^-$}
\Text(335,170)[]{$e^+$}
\ArrowLine(340,120)(360,150)
\ArrowLine(360,150)(340,180)
\Photon(360,150)(400,150){4}{6}
\Text(380,163)[]{$\gamma (Z)$}
\ArrowLine(400,150)(420,180)
\ArrowLine(420,120)(400,150)
\Text(402,132)[]{$\chi_j$}
\Text(430,180)[]{$\chi_i$}
\DashArrowLine(420,120)(420,90){4}
\ArrowLine(450,120)(420,120)
\Text(450,110)[]{$\bar{f}$}
\Text(410,90)[]{$\tilde{f}_i$}
\Text(0,45)[]{${\bf d)}$}
\Text(20,45)[]{$e^+$}
\Text(20,-30)[]{$e^-$}
\ArrowLine(55,30)(15,30)
\ArrowLine(95,30)(55,30)
\Text(75,40)[]{$\chi_j$}
\DashArrowLine(55,-20)(55,30){4}
\Text(80,5)[]{$\tilde{e}_{L,R}$($\tilde{\nu}_e)$}
\ArrowLine(15,-20)(55,-20)
\ArrowLine(55,-20)(95,-20)
\Text(95,-30)[]{$\chi_i$}
\ArrowLine(130,50)(95,30)
\DashArrowLine(95,30)(130,10){4}
\Text(135,50)[]{$\bar{f}^\prime$}
\Text(135,10)[]{$\tilde{f}_i$}
\Text(160,45)[]{${\bf d')}$}
\Text(180,45)[]{$e^+$}
\Text(180,-30)[]{$e^-$}
\ArrowLine(215,30)(175,30)
\ArrowLine(215,30)(255,30)
\Text(255,40)[]{$\chi_i$}
\DashArrowLine(215,-20)(215,30){4}
\Text(225,5)[]{$\tilde{\nu}_e$}
\ArrowLine(175,-20)(215,-20)
\ArrowLine(255,-20)(215,-20)
\Text(235,-30)[]{$\chi_j$}
\ArrowLine(290,0)(255,-20)
\DashArrowLine(255,-20)(290,-40){4}
\Text(297,0)[]{$\bar{f}^u$}
\Text(297,-40)[]{$\tilde{f}_i^d$}
\Text(310,45)[]{${\bf e)}$}
\Text(335,45)[]{$e^+$}
\Text(335,-30)[]{$e^-$}
\ArrowLine(370,30)(330,30)
\ArrowLine(410,30)(370,30)
\Text(415,40)[]{$\chi_i^0$}
\DashArrowLine(370,-20)(370,30){4}
\Text(385,5)[]{$\tilde{e}_{L,R}$}
\ArrowLine(330,-20)(370,-20)
\ArrowLine(370,-20)(395,-20)
\ArrowLine(410,-20)(395,-20)
\Text(395,-30)[]{$\chi_j^0$}
\ArrowLine(445,0)(410,-20)
\DashArrowLine(410,-20)(445,-40){4}
\Text(450,0)[]{$\bar{f}$}
\Text(450,-40)[]{$\tilde{f}_i$}
\end{picture}
\vspace*{1cm}
\caption[]{Generic Feynman diagrams contributing to the associated production
of sfermion--fermion--gaugino final states in $e^+e^-$ collisions.} 
\end{figure}
\begin{figure}[htbp]
\vspace*{1cm}
\hspace*{3cm}
\begin{picture}(1000,180)(0,0)
\Text(0,180)[]{${\bf a)}$}
\Text(15,130)[]{$e^-$}
\Text(15,170)[]{$e^+$}
\ArrowLine(20,120)(40,150)
\ArrowLine(40,150)(20,180)
\Photon(40,150)(80,150){4}{6}
\Text(60,163)[]{$\gamma, Z$}
\DashArrowLine(100,120)(80,150){4}
\DashArrowLine(80,150)(100,180){4}
\Text(82,130)[]{$\tilde{q}_j^*$}
\Text(105,180)[]{$\tilde{q}_i$}
\ArrowLine(100,120)(130,120)
\ArrowLine(100,90)(100,120)
\Text(130,110)[]{$\tilde{g}$}
\Text(90,90)[]{$\bar{q}$}
\Text(160,180)[]{${\bf b)}$}
\Text(175,130)[]{$e^-$}
\Text(175,170)[]{$e^+$}
\ArrowLine(180,120)(200,150)
\ArrowLine(200,150)(180,180)
\Photon(200,150)(240,150){4}{6}
\Text(220,163)[]{$\gamma, Z$}
\ArrowLine(260,180)(240,150)
\ArrowLine(240,150)(260,120)
\Text(244,132)[]{$q$}
\Text(265,180)[]{$\bar{q}$}
\DashArrowLine(260,120)(260,90){4}
\ArrowLine(260,120)(290,120)
\Text(290,110)[]{$\tilde{g}$}
\Text(250,90)[]{$\tilde{q}_i$}
\end{picture}
\vspace*{-3cm}
\caption[]{Feynman diagrams contributing to the associated production
of squark--quark--gluino final states in $\ee$ collisions.} 
\vspace*{-.2cm}
\end{figure}
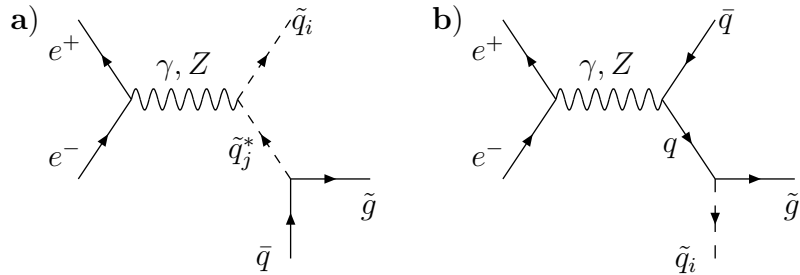

In the previous processes, all neutralino and chargino final states have to be
included, provided their masses are such than $m_{\chi_i} \leq \sqrt{s}-
m_{\tilde{f}}-m_f$, i.e. when the phase space is large enough for the
associated production to take place. For the exchanged gauginos in
Figs.~1c--1e, all states should in principle be included. However, if at the
same time $\sqrt{s} \geq m_{\chi_i}+ m_{\chi_j}$ and $m_{\chi_i} \geq
m_{\tilde{f}}+m_f$, both the two--body production process $\ee \to \chi_i
\chi_j$ and the two--body decay $\chi_i \to \tilde{f} f$ can occur, and the
total cross section would be then simply the cross section for the two--body
production process times the branching ratio for the decay channel. Since this
situation has been already analyzed in several places, we will not discussed it
further here. \s

The channels contributing to the various processes discussed above are 
summarized in Table 1, following the labels of Fig.~1. In the case of the 
associated production of sfermions with charginos, there is an overall minus 
signal difference between the contributions of the the $s$--channel gauge boson
exchanges (Fig.~1c) in the case of isospin $+1/2$ and $-1/2$ because of the 
charge of the chargino. In addition, diagram Fig.~1d$^\prime$ contributes only 
to $\tilde{d}_L (\tilde{l}_L) \bar{u}(\bar{\nu}_l)  \chi_i^+$ final states,  
while diagram Fig.~1d contributes to both $\tilde{u}_L(\tilde{\nu}_l) \bar{d}
(\bar{l}) \chi_i^-$ and $\bar{f} \tilde{f} \chi^0$ final states; with the 
conventions adopted, the common structure of the matrix elements for the latter
two processes differs only by a relative sign. \s


\vskip 20pt
\begin{center}
\renewcommand{\arraystretch}{1.5}
\begin{tabular}{|c||c|c|c|c|c|c|c|c|c|}
\hline
Final states &  $a_\gamma$ & $b_\gamma$ & $c_\gamma$ & $a_Z$ & $b_Z$ & $c_Z$ 
& $d$ & $d^\prime$ & $e$ \\
\hline
\hline
$\bar{f} \tilde{f} \chi^0$ & $\surd$ & $\surd$ &  & $\surd$ & $\surd$ & $\surd$
& $\surd$ & & $\surd$ \\
\hline
$\bar{\nu} \tilde{\nu} \chi^0$ &  &  &  & $\surd$ & $\surd$ & $\surd$
 & $\surd$ & & $\surd$ \\
\hline
$\bar{f}_d \tilde{f}_u \chi^-$ &  & $\surd$ &  $\surd$ & $\surd$ & 
$\surd$ & $\surd$ & $\surd$ & & \\
\hline
$\bar{f}_u \tilde{f}_d \chi^+$ & $\surd$ & & $\surd$ & $\surd$ & 
$\surd$ & $\surd$ & & $\surd$ & \\
\hline
$\bar{q} \tilde{q} \tilde{g}$ & $\surd$ & $\surd$ & & $\surd$ & 
$\surd$ & & & & \\
\hline
\end{tabular}
\vskip 10pt
Table 1: Contributing diagrams to the various final states.
\end{center}

\vskip 2pt
The relative signs among different diagrams, when contributing to the same
final state, arising out of the anticommuting nature of the fermionic fields 
(Wick's theorem) are summarized in Table 2 below.  \s

\vskip 10pt
\begin{center}
\renewcommand{\arraystretch}{1.5}
\begin{tabular}{|c||c|c|c|c|c|c|c|c|c|}
\hline
Diagrams &  $a_\gamma$ & $b_\gamma$ & $c_\gamma$ & 
$a_Z$ & $b_Z$ & $c_Z$ & $d$ & $d^\prime$ & $e$ \\
\hline
\hline
Relative Signs& + & + & + & + & + & + & $-$ & + & $-$ \\
\hline
\end{tabular}
\vskip 10pt
Table 2: Relative signs among different contributing diagrams.
\end{center}

Finally, we note that the production of the charged conjugate states has also 
to be taken into account. Due to CP--invariance, these cross sections are the 
same as for the corresponding previous ones, which have thus to be multiplied 
just by a factor of two. 

\subsubsection*{2.2 The amplitudes of the various contributions}

\noindent
The generic process considered here, including the momenta  $p_i$'s of all 
particles, is
\beq
e^+(p_1) \; e^-(p_2) \to \tilde{f}_j (p_3) \; \chi_i(p_4) \; 
\bar{f}^\prime(p_5) 
\eeq
where $\chi$ stands for an electroweak gaugino $\chi_{1,2}^0$ or $\chi_1^\pm$. 
The amplitudes of the various channels, following the labels of Fig.~1, are 
given in a covariant gauge by:
\beq
M_a^{\gamma} &=& -\frac{e^3 Q_f}{\stw} \sum_n 
\frac{[\bar{v}_e(p_1) \, (\pn3-\pn4-\pn5) \, u_e(p_2) ] \;
[\bar{u}_{\chi_i}(p_4) \, G^{Ain}_{a \gamma} \, v_f(p_5)]} 
{(p_1+p_2)^2 \, \{(p_4+p_5)^2 - m_{\tilde{f}_n}^2\} } \non \\ \non \\
M_b^{\gamma} &=& -\frac{e^3 Q_f}{\stw} 
\frac{[\bar{v}_e(p_1) \, \gamma^{\alpha} \, u_e(p_2) ] \;
[\bar{u}_{\chi{_i}}(p_4) \; G^{Ain}_{b \gamma} \, (\pn3+\pn4+m_f) \, 
\gamma_{\alpha} \, v_f(p_5)]}{(p_1+p_2)^2 \, \{(p_3+p_4)^2-m_f^2\} } \non \\ 
\non \\ 
M_c^{\gamma} &=&  \frac{e^3 Q_\chi}{\stw}
\sum_j \frac{[\bar{v}_e(p_1) \, \gamma^{\alpha} \, u_e(p_2) ] \;
[\bar{u}_{\charpmi}(p_4) \, \gamma_{\alpha} \, (\pn3+\pn5-m_{\charpmj})
\, G^{Ajn}_{c \gamma} \, v_f(p_5)]}
{(p_1+p_2)^2 \, 
\{(p_3+p_5)^2 - m_{\charpmj}^2 \}} \non \\ \non \\ 
M_a^Z &=& \frac{e^3 a_{Z\tilde{f}_i\tilde{f}_j}}{\stwn3 \ctwn2}
\sum_n
\frac{
[\bar{v}_e(p_1) \, (\pn3-\pn4-\pn5) \, (c_L^e P_L+c_R^e P_R) \; u_e(p_2) ] \;
\; [\bar{u}_{\chi_i}(p_4) \; G^{Ain}_{a Z} \, v_f(p_5)]}
{\{(p_1+p_2)^2 -m_Z^2\} \, \{(p_4+p_5)^2 - m_{\tilde{f}_n}^2 \} } 
\non \\ \non \\ 
%
M_b^Z &=& 
\frac{e^3/ (\stwn3 \ctwn2)}
{\{(p_1+p_2)^2-m_Z^2 \} \, \{(p_3+p_4)^2 - m_f^2\}} \;
[ \bar{v}_e(p_1) \, \gamma^\alpha \, (c_L^e P_L+c_R^e P_R) \, u_e(p_2) ]
\non \\ && \times 
[\bar{u}_{\chi{_i}}(p_4) \; G^{Ain}_{b Z} \, (\pn3+\pn4+m_f) \, \gamma_\alpha
\, (c_L^f P_L+c_R^f P_R) \, v_f(p_5)] \non \\ \non \\ 
%
M_c^Z &=&  (-1)^{Q_\chi} \frac{e^3}{\stwn3 \ctwn2} \sum_j 
\frac{[\bar{v}_e(p_1) \, \gamma^\alpha \,(c_L^e P_L+c_R^e P_R) \, u_e(p_2) ]}
{\{(p_1+p_2)^2-m_Z^2 \} \, \{(p_3+p_5)^2 - m_{\chi_j}^2\} } \non \\
&& \times [\bar{u}_{\chi{_i}}(p_4) \, \gamma_\alpha \; g_{Z\chi_i \chi_j} \; 
(\pn3+\pn5-m_{\chi_j}) \; G^{Ajn}_{c Z} \, v_f(p_5)] \non \\ \non \\ 
%
M_d &=& \frac{e^3}{\stwn3} \sum_{\tilde{e}_{L,R}} \sum_j \frac{
[\bar{v}_e(p_1)\, G^{Bj}_{dL(R)}\, (\pn3+\pn5-m_{\chi_j}) \; G^{A_2jn}_{d} 
\, v_f(p_5)]
\; [\bar{u}_{\chi{_i}}(p_4) \; G^{A_1i}_{dL(R)} \; u_e(p_2)]}
{ \{(p_3+p_5)^2 - m_{\chi_j}^2 \} \; 
\{(p_2-p_4)^2 - m_{(\tilde{e}_{(L,R)},\tilde{\nu}_{e})}^2 \}} \non \\ \non \\
%
M_{d^\prime} &=& - \frac{e^3}{\stwn3} \sum_j \frac{
[\bar{v}_e(p_1)\, G^{Bi}_{d^\prime \tilde{\nu}} \; v_{\chi_i^\pm}(p_4)] \; 
[\bar{u}_f(p_5) \, G^{A_2jn}_{d^\prime} \, (\pn3+\pn5+m_{\chi_j^\pm}) \, 
G^{A_1j}_{d^\prime \tilde{\nu}} \, u_e(p_2)]}
{ \{(p_3+p_5)^2 - m_{\chi^\pm_j}^2 \} \; 
\{(p_1-p_4)^2 - m_{\tilde{\nu}_{e}}^2 \}} \non \\ \non \\ 
%
M_e &=& \frac{e^3}{\stwn3} \sum_{\tilde{e}_{L,R}} \sum_j 
\frac{[\bar{v}_e(p_1)\, G^{Bi}_{e L(R)}\, v_{\chi_i^0}(p_4)]
[\bar{u}_f(p_5) \, G^{A_2jn}_e (\pn3+\pn5+m_{\chi_j^0}) 
G^{A_1j}_{e L(R)}\, u_e(p_2)] }
{ \{(p_3+p_5)^2 - m_{\chi_j^0}^2 \}  
\{(p_1-p_4)^2 - m_{\tilde{e}_{(L,R)}}^2 \}} 
\eeq

The left-- and right--handed projectors are $P_{L,R}=\frac{1}{2}(1\mp
\gamma_5)$. The constants and the couplings used in writing down these
amplitudes are described and defined below, in terms of the electromagnetic
coupling constant $e^2 = 4\pi \alpha$, the sine and the cosine of the Weinberg
angle, $\stw=\sin\theta_W$ and $\ctw=\cos\theta_W$, $T_{3,\tilde{f}_k(f)}$ and
$Q_{\tilde{f}_k(f)}$, which respectively, are the third component of the weak
isospin and the charge of the $k$-th chiral sfermion (fermion). Note that in
the case of gluino final states, only a subset of diagrams will contribute and
the QCD factors have to be included. \s

\nn {\it Fermion and sfermion couplings with gauge bosons:} 
\beq
c_L^f &=& T_{3,f} - Q_f \stwn2 \qquad \quad , \quad
c_R^f = - Q_f \stwn2 \non \\ 
a_{Z \tilde{f}_i \tilde{f}_j} &=& \sum_{k=L,R} a_k M_{ik} M_{jk} \qquad 
{\rm with} \quad a_k = T_{3,\tilde{f}_k} - Q_{\tilde{f}_k} \, \stwn2 
\eeq
The $2\times2$ matrix $M$ is the orthogonal rotation matrix connecting the
mass ($i,j$) and the chiral ($L,R$) eigenstates of the sfermions and defined as
\begin{eqnarray}
\left(  \begin{array}{c} \tilde{f}_1 \\ \tilde{f}_2 \end{array} \right) 
= \left( \begin{array}{cc} M_{1L} & M_{1R} \\ M_{2L} & M_{2R}\end{array} \right)
\; \left( \begin{array}{c} \tilde{f}_L \\ \tilde{f}_R \end{array} \right)   
= \left( \begin{array}{cc} \cos\theta_{\tilde{f}} & \sin\theta_{\tilde{f}} \\
-\sin\theta_{\tilde{f}} & \cos\theta_{\tilde{f}} \end{array} \right) \;
\left( \begin{array}{c} \tilde{f}_L \\ \tilde{f}_R \end{array} \right) 
\end{eqnarray}
where, $\theta_{\tilde{f}}$ is the mixing angle.
For the first two generations of sfermions
the mixing effect can reasonably be neglected  (i.e. $\theta_{\tilde{f}}=0$)
so that $M_{ln}=0$  if $l \neq n$. 
Particularly, for sneutrinos, only $M_{1L} \neq 0$. \s

\nn {\it Gaugino-gaugino-$Z$ boson couplings:} \s

We followed Figs.~75(b,c,d) and eqs.~(C88) of Haber and Kane \cite{HaberKane}
for the $\gamma \chi^+_i \chi^-_j$ (with $Q_{\chi_i^{\pm,0}} = \pm 1, 0$ being
the charge of the on--shell gaugino) and the  $Z \chi_i \chi_j$ couplings, the 
latter being defined as $g_{Z\chi_i\chi_j} = O^L_{ij} P_L + O^R_{ij} P_R$, 
with: 
\begin{eqnarray}
{\rm For~charginos} &:&   \begin{array}{l}
O^L_{ij} = -V_{i1} V_{j1} - \frac{1}{2} V_{i2} V_{j2} + \delta_{ij} s_W^2 \\ \\
O^R_{ij} = -U_{i1} U_{j1} - \frac{1}{2} U_{i2} U_{j2} + \delta_{ij} s_W^2 
\end{array} \non \\
{\rm For~neutralinos} &:&  \ \ O^L_{ij} = - O^R_{ij} =
-\frac{1}{2} N_{i3} N_{j3} + \frac{1}{2} N_{i4} N_{j4} 
\end{eqnarray}

\vskip 10pt
\nn {\it Fermion-sfermion-gaugino couplings:} \s

For the couplings of charginos and neutralinos to fermions and scalar fermions
we follow Figs.~22,23 and 24 of Haber and Gunion \cite{HaberGunion}. In the
amplitudes presented in Eq.~(5), the couplings $G$ absorb the sign on imaginary
$i$'s as shown against the vertices in the figures mentioned above. The
subscripts of $G$ indicate the diagram and, if appropriate, the propagator
gauge boson.  On the other hand the superscripts ($i$ or $j$) indicate whether
this coupling is arising at a vertex with an outgoing ($i$) or a propagator
($j$) gaugino.  A superscript $n$ in appropriate cases indicates the eigenstate
of the scalar propagator.  Couplings with superscript $A$ and $B$ are related
by hermitian conjugation at the Lagrangian level, while involving the same set
of fields in respective cases.  There are two $A$--type couplings for the
$t$--channel processes in diagrams $(1d)$, $(1d')$ and $1e$. $A_1$ is
associated with the vertex with the incoming $e^-$ while $A_2$ is associated
with the vertex where the outgoing sfermion appears. Note that while $A$--type
couplings appear in both $s$-- and $t$--channel diagrams involving both final
state gaugino and sfermion, $B$--type couplings only appear in $t$--channel and
associate with the incoming $e^\pm$ coupled either to a final state gaugino or
one in the propagators. Also, note that in the most general case all the
indices on the couplings $G^{A,B}$ are nontrivial in the sense that couplings
of a particular type with a generic structure may assume different values not
only from diagram to diagram but also within a diagram and hence need special
care while defining them. \s

The $q\tilde{q}\tilde{g}$ coupling follows from Eq.~(C89) of Haber and Kane
\cite{HaberKane}.  Special care should be taken in reading out the $s$-channel
$\gamma$ and $Z$-mediated amplitudes of diagrams Figs.~2a and 2b when a single
factor of $e/\stw$ should be replaced by the strong coupling constant $g_s$. In
addition one has to include in the amplitudes, the Gell-Mann matrices which
appear in the squark--quark--gluino couplings. \s

In all the couplings defined above we have taken the neutralino-mixing matrix 
$N$ in the $B-W^3$ basis, the neutralino-mixing matrix $N'$ in the
$\tilde{\gamma}-\tilde Z$ basis, the two chargino mixing matrices $U$ and $V$,
to be real as is appropriate in an analysis that conserves CP. Following are 
the couplings in details.

\vskip 10pt
\nn $A$--type couplings in $s$--channel, where $s=a,b,c$ and $V$ stands for 
$\gamma$ and $Z$ boson:
\beq
G^{Ai(j)n}_{sV} &=& C^A_1 (A_{i(j)1} P_L + A_{i(j)2} P_R) M_{nL}
                    + C^A_2 (A_{i(j)3} P_L + A_{i(j)4} P_R) M_{nR} 
\eeq
$A$--type couplings in $t$--channel, where $t=d,d',e$:
\beq
G^{A_1i(j)}_{t(L,R,\tilde{\nu})} &=&C^{A}_{L(R)} (A_{i(j)1(L,R,
\tilde{\nu})} P_L + A_{i(j)2(L,R,\tilde{\nu})} P_R) \non \\
G^{A_2jn}_{t} &=& G^{Ajn}_{cZ} 
\eeq
$B$--type couplings which appear only in $t$--channel:
\beq
G^{Bi(j)}_{L(R)} &=& C^B_{L(R)} (B_{i(j)1L(R)} P_R + B_{i(j)2L(R)} P_L) \non \\
G^{Bi}_{\tilde\nu} &=& C^B_1 (B_{i1} P_R + B_{i2} P_L) 
\eeq
The generic structures of the terms on the right hand side of the above 
equations are defined below in reference to three final states broadly 
categorized as final states with neutralino, chargino and that with gluino.\s

\nn For final states with neutralinos, one has:
\beq
C^A_1 &=& C^A_2 = C^{A}_{L(R)} = C^{B}_{L(R)} = -1 \non \\ \non
A_{i(j)1L} &=& B_{i(j)1L} =
-\sqrt{2} [\stw N^\prime_{i(j)1} + (\frac{1}{2}-\stwn2) 
N^\prime_{i(j)2})] \non \\
A_{i(j)2R} &=& B_{i(j)2R} = 
\sqrt{2} (\stw N^\prime_{i(j)1} - \stwn2 N^\prime_{i(j)2}) \non \\
A_{i(j)2L} &=&  A_{i(j)1R} = B_{i(j)2L} =  B_{i(j)1R} = 0 
\eeq
For charged sleptons and $d$-type squarks (sneutrinos and $u$-type squarks)
in the final state:
\beq
A_{i(j)1} &=& \sqrt{2} [\stw Q_f N^\prime_{i(j)1} + (T_{3,f}-Q_f \stwn2) 
N^\prime_{i(j)2}] \non \\
A_{i(j)2} &=& A_{i(j)3} = \frac{m_f}{\sqrt{2} m_W} r^f
\qquad {\rm with} \;\; r^d(r^u) =\frac{N_{i(j)3}}{\cos\beta} 
\left (\frac{N_{i(j)4}}{\sin\beta}\right) \non \\
A_{i(j)4} &=& -\sqrt{2} (\stw Q_f N^\prime_{i(j)1} 
-\stwn2 Q_f N^\prime_{i(j)2})  
\eeq
For final states with charginos, one has:
\beq
A_{i(j)1L} &=& V_{i(j)1} \ \ , \qquad A_{i(j)2L} = A_{i(j)1R} = A_{i(j)2R} = 
0 \non \\
B_{jl(L,R)} &=& A_{jl(L,R)} \qquad  {\rm with}~ l=1,2 
\eeq
For charged sleptons or $d$-type squarks (sneutrinos or $u$-type squarks) 
in the final state:
\beq
C^{A}_L &=& -C^{B}_L = +1  \qquad C^{A}_R =  C^{B}_R = 0 \qquad 
C^A_1 = -C^A_2 = -1 (+1) \non \\
C^B_1 &=& -C^B_2 = -1 \qquad ({\rm appears \; only \; for \; the \; 
case \; of \;} \tilde{l}^+ \; {\rm and} \; \tilde{d})  \non \\ \non \\
A_{i(j)1} &=& U_{i(j)1} \; (V_{i(j)1}) \ \ , \qquad A_{i(j)4} = 0 \ \non \\
A_{i(j)2} &=& -\frac{m_f}{\sqrt{2} m_W} r_1^f \quad {\rm with} \quad 
r_1^d (r_1^u) = \frac{V_{i(j)2}}{\sin\beta} \; \left(\frac{U_{i(j)2}}{\cos
\beta} \right) \non \\
A_{i(j)3} &=& -\frac{m_{f^\prime}}{\sqrt{2} m_W} r_2^f \quad {\rm with} \quad 
r_2^d (r_2^u) = \frac{U_{i(j)2}}{\cos\beta} \; \left(\frac{V_{i(j)2}}
{\sin\beta} \right)  \\ \non \\
B_{jl} &=& A_{jl} \quad  {\rm with } \ l=1 \ldots 4 \non \\ 
B_{i1} &=& V_{i1} \ \ , \quad 
B_{i2} = B_{i3} = B_{i4} =  0  \qquad ({\rm appears \; only \; for \; the \; 
case \; of \;} \tilde{l}^+ \; {\rm and} \; \tilde{d}) \non 
\eeq
Finally, for gluino final states, we have: 
\beq
C^A_1 = -C^A_2 = -1 \ , \quad A^I_{i1} = A^I_{i4} = \sqrt{2} \ , \quad
A^I_{i2} = A^I_{i3} = 0 
\eeq

\vskip 10pt
In our analysis, we will impose kinematic constraints prohibiting on--shell
limits for the (sfermions and/or gauginos) propagators that would have led to
resonances when the sparticle widths, i.e. the Breit-Wigner form of the
sparticle propagators, are not explicitly used. This is quite justified as
long as we are interested in studying situations reasonably away from such
thresholds. But including sparticle widths in the propagators is a must when we
are rather close to the thresholds; see for instance Ref.~\cite{Peter}. We will
demonstrate this effect separately in the study of pair-production of muons 
along with two neutralinos in a 4--body final state in $\ee$ collisions with 
an emphasis on the role of finite widths of sparticles close to such 
thresholds.

\subsubsection*{2.3 The differential cross sections} 

The analytical expressions of the amplitudes squared of the various processes
are quite lengthy and not very telling\footnote{The Fortran code which
calculates the total cross section and includes all the amplitudes squared and
the interferences can be obtained from {\tt datta@lpm.univ-montp2.fr}.}. In the
Appendix, we will simply give the amplitude squared of the sum of the three
diagrams Fig.~1a--c, the contribution of which is the dominant one as will
discussed later and explicitly shown in section 4. In principle, close to the
$\sqrt{s} \sim 2 m_{\tilde{f}}$ threshold, diagrams (1a) with an (almost)
resonant intermediate sfermion should give the dominant contribution. Below
this threshold, an important contribution will come from diagram (1b) with a
virtual fermion, and in the case of associated production of
sfermion--fermion--chargino final states, from diagram (1c).  However, these
diagrams are not gauge invariant by themselves, and only the sum of the three
contributions (1a+1b+1c) is gauge invariant\footnote{In a covariant gauge, the
gauge dependent term is present only in the longitudinal component of the
photon and $Z$--boson propagators and should therefore give a vanishing
contribution when it is attached to the initial leptonic tensor in the limit of
zero electron mass. However, in other (non--covariant) gauges such as the axial
gauge, the gauge dependent terms are explicitly non--zero in the individual
contributions of diagrams (1a), (1b) and (1c) and only the sum of the three
contributions is free of this gauge dependent terms. In particular, diagram
(1c) which, for associated smuon and LSP production as will be discussed later,
should give a contribution that is proportional to the small ${\cal
O}(M_2^2/\mu^2)$ $Z$--$\chi$--$\chi$ coupling in the gaugino region, can be
neglected only when the calculation is performed in a covariant gauge. This is
not automatically the case in a general gauge and to be consistent, we will
include also this contribution in the analytical expression of the
differential cross sections. We thank Ayres Freitas and Peter Zerwas for a
discussion on this point.}.  We will further use the simplification that the
off--shell sfermion is of the same species as the real one, which is a very good
approximation in the case of first and second generation sfermions where the
mixing is very small and the $Z$--boson does not couple to different chiral
sfermions. In the case of third generation sfermions, the mass splitting
between the sfermion eigenstates is large if the mixing is strong and the
contribution of the heavier sfermion is suppressed by its large virtuality. 
So here also, we carry on with the lighter species in the final state only. 
[Note that the case of first and second generation squarks is similar to the 
smuon case].
\s 



\newpage

\subsection*{3. The associated production cross sections} 

\subsubsection*{3.1 The weight of the various contributions}

Before presenting our numerical results, let us discuss the contributions from
diagrams Fig.~1a-e to the production cross sections, focusing on the two
extreme cases where the higgsino mass parameter $\mu$ is much smaller or much
larger than the gaugino mass parameters $M_2$. All through this analysis, we 
also presume the unification of gaugino masses at the GUT scale, leading to the
relations $M_1=\frac{5}{3} \tan^2 \theta_W M_2 \simeq \frac{1}{2} M_2$ 
and $m_{\tilde{g}} \simeq M_3 \simeq 3.5 M_2$ at the weak scale.\s

If $|\mu| \gg M_{1,2}$, the lightest chargino $\chi_1^\pm$ and the two lighter
neutralinos $\chi_{1,2}^0$ are gaugino--like states with masses $m_{\chi_{1}^0}
\simeq M_1$, $m_{\chi_{2}^0} \simeq m_{\chi_1^\pm} \simeq M_2 \simeq 2M_1$,
while heavier charginos and neutralinos are higgsino--like states with masses
$m_{\chi_{3}^0} \simeq m_{\chi_{4}^0} \simeq m_{\chi_2^\pm} \simeq |\mu|$. The
couplings of the gauginos to electron--selectron or electron--sneutrino states
are maximal, while the couplings of the higgsinos to sfermions and massless
fermions vanish. [Note that in this case, the couplings of right--handed
sfermions to the gauginos $\chi_1^\pm$ and $\chi_2^0$ also vanish].  In this
case, only the contributions of the diagrams with light gauginos in the final
state, and light gauginos exchanged in the diagrams Fig.~1c--1e,  have to be
considered. [Contributions with heavier higgsino final states, in the case of
smuon and sneutrino production, are suppressed, in addition to the smaller
phase space, by one, two or even three powers of the small
higgsino--lepton--slepton coupling.] In the case of $\tau$ sleptons, the mixing
can be very strong for large values of $\tan \beta$ and $\mu$ leading to
non--negligible higgsino--$\tilde{\tau}$--$\tau$ or higgsino--$\tilde{\tau}$--$
\nu_\tau$ couplings. In this case not only the $s$--channel diagrams (1a) and
(1b) would contribute but also diagram (1c) for higgsino final states, although
it might be suppressed by phase space. The situation for third generation
$\tilde{t}$ and $\tilde{b}$ squarks is similar to that of the $\tilde{\tau}$
slepton. \s

In the opposite situation, $|\mu|\ll M_{1,2}$, the lighter chargino
$\chi_1^\pm$and neutralinos $\chi_{1,2}^0$ will be higgsino like and degenerate
in mass.  Their couplings to selectrons and sneutrinos will be very tiny and
therefore the diagrams (1d, 1d') and (1e) will give very small contributions
with light chargino and neutralinos in the final or intermediate states.  In
this case, processes involving heavier neutralino and chargino final [or
intermediate in the case of Fig.~1d--1e] states may significantly contribute to
the cross sections, but since these states are rather heavy, the processes will
be kinematically disfavored [in the case of intermediate gaugino states, they
will be suppressed  by the larger virtuality of the particles too]. Again
diagram (1c) will give a very small contribution, but this time, because of the
tiny final higgsino coupling to sfermion--fermion pair [except in the case of
the top squarks because of the large value of $m_t$, and $\tilde{\tau},
\tilde{b}$ sfermions for very large $\tb$ values]. \s

In the mixed gaugino--higgsino region, i.e. for $|\mu| \sim M_{2}$, and for
associated production of third generation $\tilde{\tau}$ leptons and
$\tilde{t}, \tilde{b}$ squarks with chargino and neutralinos, the situation is
more complicated [as is well known] because of the large mixing in both the
sfermion and gaugino sectors and all types of diagrams involving all gaugino
states have to be taken into account. However, for a given final state, the
cross section will be in general smaller than the one in the pure gaugino case,
despite a more favorable phase space [since all neutralinos and charginos will
have comparable masses in this case]. This is because  the couplings of
chargino and neutralinos to sfermions [in particular to sleptons] are reduced
compared to the gaugino case.  \s

Thus, we will concentrate in our analysis on the more favorable case where the
higgsino mass parameter $\mu$ is much larger than $M_{1,2}$, i.e. on the
gaugino LSP case, although we will give some illustrations in the other cases.

\subsubsection*{3.2 Numerical analysis} 

In this section we will illustrate the magnitude of the total production cross
sections for the various processes discussed above. We will set for
definiteness $\tb=30$, and in the main part of the analysis, we will choose a
large value of the higgsino mass parameter, $\mu=500$ GeV, and a low value of
the wino mass parameter, $M_2\sim 100$ GeV, leading to gaugino--like lighter
chargino and neutralinos with approximate masses $m_{\chi_1^\pm} \sim
m_{\chi_2^0} \sim 2 m_{\chi_1^0} \sim 100$ GeV, i.e.  values close to the
experimental LEP2 bounds \cite{LEP2}. The large $\tb$ and $\mu$ values will
imply a large mixing in the $\tilde{\tau}$ and $\tilde{b}$ sectors, leading to
relatively light states compared to $\tilde{\mu}$ and $\tilde{q}$,
respectively. For $\tilde{\tau}$ sleptons, the no--mixing case is realized from
the $\tilde{\mu}$ case, where the large value of $\tb$ and $\mu$ do not have a
substantial effect since the smuon mass $m_\mu$ is too small. The situation for
the two sneutrinos $\tilde{\nu}_\tau$ and $\tilde{\nu}_\mu$ will be of course
identical. \s

For the soft--SUSY breaking scalar fermion masses at low energies, we  use
a universal value $m_{\tilde{f}_L}= m_{\tilde{f}_R}$. To obtain the masses
$m_{\tilde{f}_1}$ and $m_{\tilde{f}_2}$ one has to include the ``D--terms" but
since their contributions are rather small for $m_{\tilde{f}_{L,R}}$ of the
order of a few hundred GeV, the mass eigenstates $\tilde{f}_1$ and
$\tilde{f}_2$ will be almost identical to the current eigenstates and mass
degenerate. The degeneracy is lifted in the case of third generation charged
sfermions by the mixing which is proportional to $ m_f (A_f- \mu \tan \beta)$
for the isospin $-\frac{1}{2}$ $\tilde{b}$ and $\tilde{\tau}$ states and $m_t
(A_t-\mu/\tb)$ for the isospin $+\frac{1}{2}$ top squarks. In the analysis, we
will set the trilinear couplings $A_f$, which has no major effect in this
context, to zero.  \s

Note that we have thoroughly cross checked our results against the
corresponding ones from the program {\tt CompHEP} \cite{CompHep} in the cases
of $\tilde{\mu}_{L,R}, \tilde{\nu}_\mu$ and found very good agreement.
\s 

Fig~3 shows the cross sections for the production of smuons and their sneutrino
partners in association with the lightest neutralino (3a) and the lightest
chargino (3c) as well as the production of the lightest $\tilde{\tau}_1$
slepton in association with the lightest chargino and neutralino (3b). The
cross sections are shown for a c.m. energy of 500 GeV as a function of 
slepton mass\footnote{In all the plots, we start the variation of the sfermion
masses at approximately 2 (4) GeV above the kinematical threshold for
$\sqrt{s}=500$ GeV (1 TeV). These values typically correspond to the total
decay width of the produced sfermion with masses of 250 (500) GeV. Close to
the kinematical threshold, a more dedicated analysis is required as will be
discussed in section 4.}, for the SUSY parameters discussed in the previous
section. Shown are the summed up cross sections for the processes $\ee \to 
\bar{\ell} \tilde{\ell} \chi$ and the charge conjugate process $\ee \to \ell
\tilde{\ell}^* \chi$, which means that if the production of only one
state is to be considered, the cross sections should be divided by a factor of
two. \s

In Fig.~3a, one sees that the cross section for the production of the LSP with
the right--handed smuon $\tilde{\mu}_R$ is larger than the one for the
production with the left--handed smuon $\tilde{\mu}_L$ and its partner
sneutrino $\tilde{\nu}_\mu$, in particular near the kinematical threshold
$\sqrt{s}=2 m_{\tilde{\ell}}$. For instance, the cross section for the process
$\ee \to \mu \tilde{\mu}_R \chi_1^0$ is close to 0.05 fb for
$m_{\tilde{\mu}_R}=260$ GeV. This means that with the very high integrated
luminosities, $\int {\cal L} \sim 500$ fb$^{-1}$, expected at future linear
colliders \cite{TESLA}, 25 of such events can be collected within one year,
opening the possibility of discovering $\tilde{\mu}_R$ for masses of 10 GeV
above the threshold. The cross sections decrease quickly with larger
$m_{\tilde{\ell}}$, dropping to below 0.01 fb for all three processes [i.e. 5
events for the above luminosity] for masses close to 300 GeV. \s

The production cross sections of the $\tilde{\mu}_L$ and its partner sneutrino
in association with the lightest chargino are shown in Fig.~3b
[$\tilde{\mu}_R$ does not couple to the $\chi_1^\pm$ states in this regime],
and as can be seen, they are approximately a factor of two larger than the
corresponding production cross sections in association with the LSP, a mere
consequence of the stronger charged current couplings compared to the neutral
current ones. \s

The cross sections for the associated production of the lightest
$\tilde{\tau}$ lepton with the lightest charginos and neutralinos, Fig.~3c, are
similar to those of $\tilde{\mu}_L$ in spite of sfermion mixing. As
mentioned earlier, the cross sections for the associated production of both
types of sneutrinos, $\tilde{\nu}_\mu$ and $\tilde{\nu}_\tau$, with charginos
is the same.  \s

\begin{figure}[htbp]
\begin{center}
\vskip-4cm
\hskip-1cm\centerline{\epsfig{file=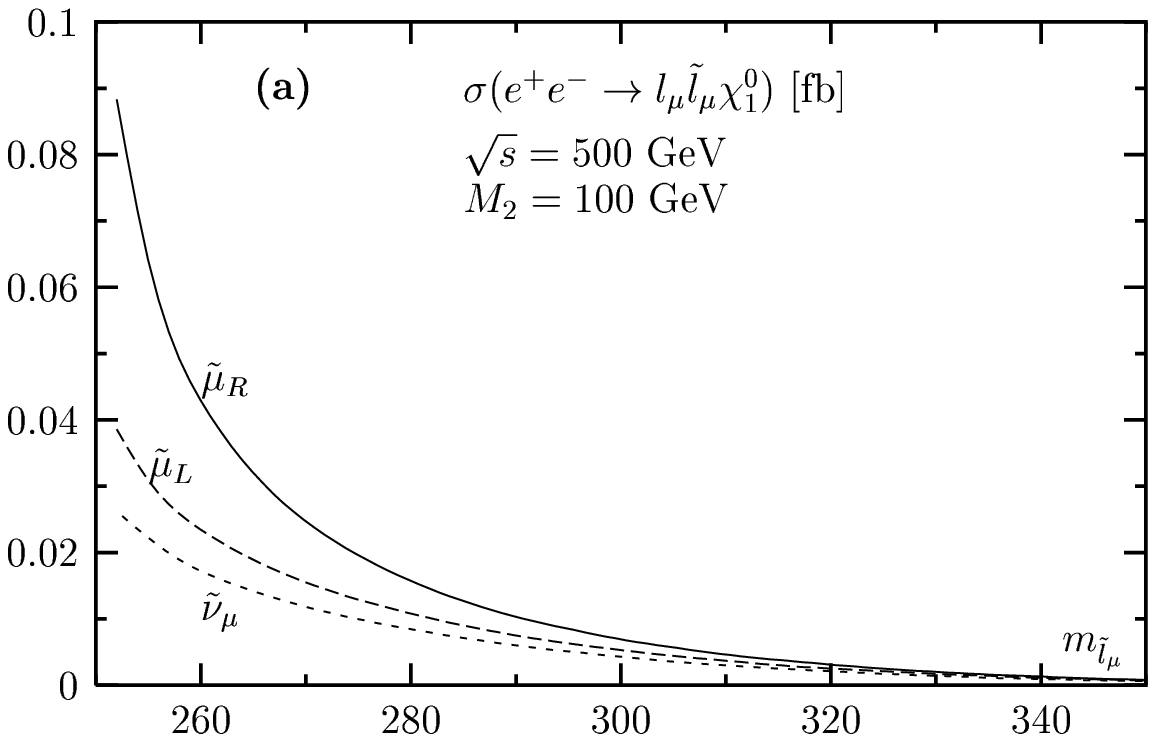,width=17cm}}
\vskip-17.5cm
\hskip-1cm\centerline{\epsfig{file=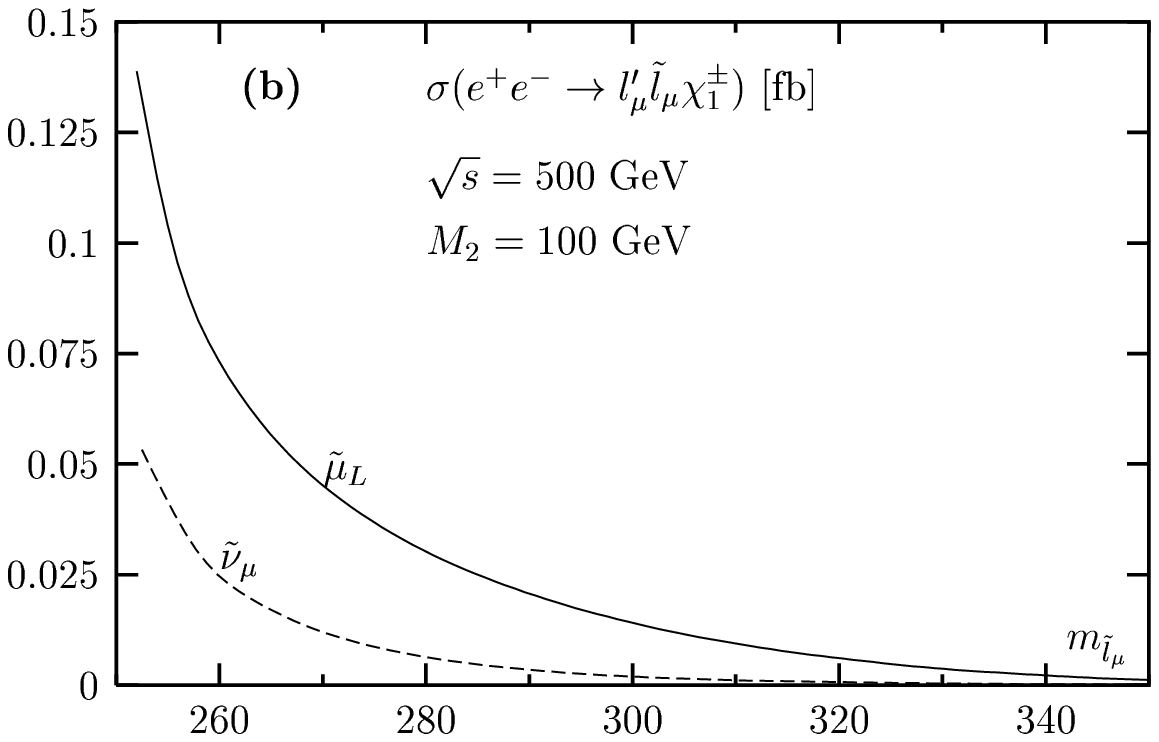,width=17cm}}
\vskip-17.5cm
\hskip-1cm\centerline{\epsfig{file=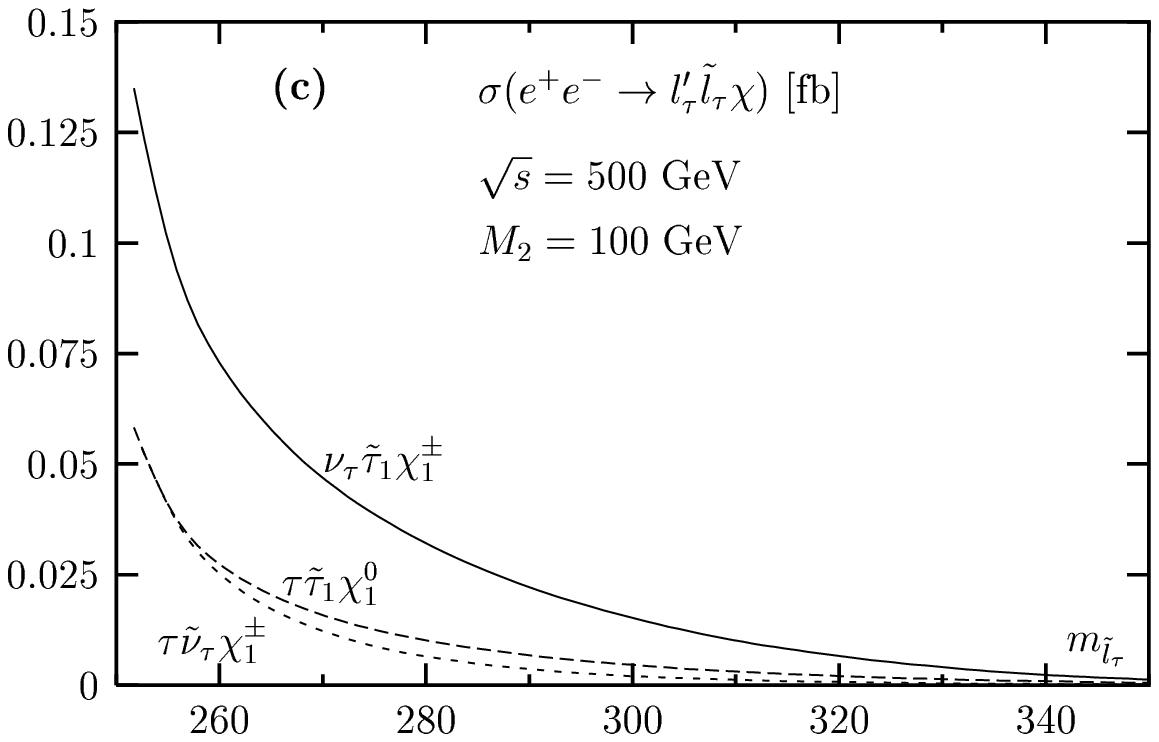,width=17cm}}
\vskip-14.cm
\caption{Associated production cross sections for sleptons and gauginos 
in $\ee$ collisions for a c.m. energy $\sqrt{s}=500$ GeV as a function 
of the slepton masses. For the SUSY parameters, we have set $m_{\tilde{\ell}_L}
= m_{\tilde{\ell}_R}, M_2=100$ GeV, $\mu=500$ GeV and $\tb=30$.}  
\end{center}
\end{figure}

\begin{figure}[htbp]
\begin{center}
\vskip-4cm
\hskip-1cm\centerline{\epsfig{file=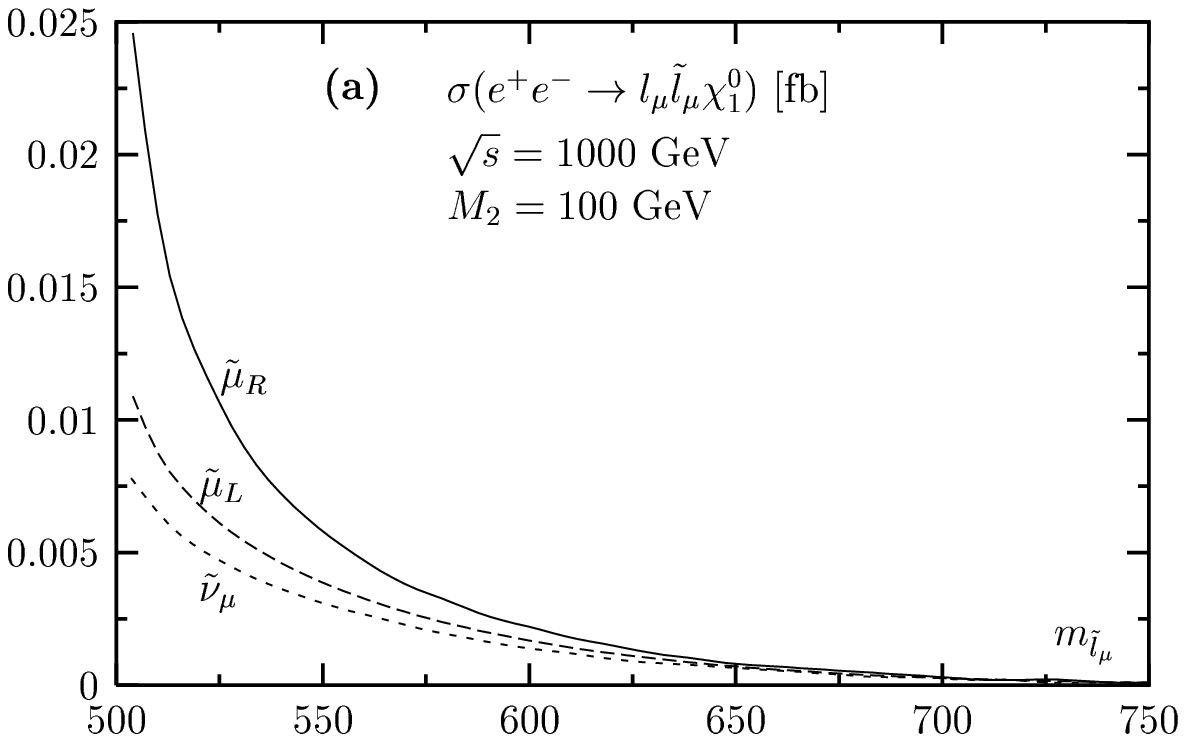,width=17cm}}
\vskip-17.5cm
\hskip-1cm\centerline{\epsfig{file=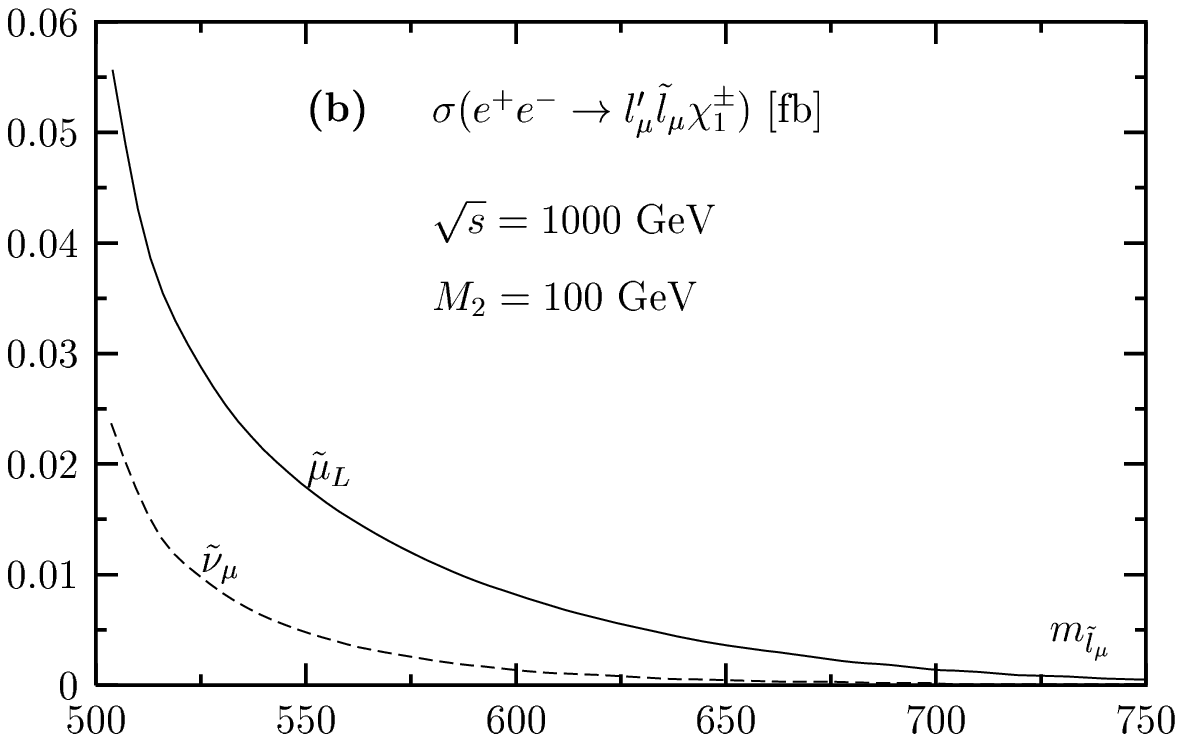,width=17cm}}
\vskip-17.5cm
\hskip-1cm\centerline{\epsfig{file=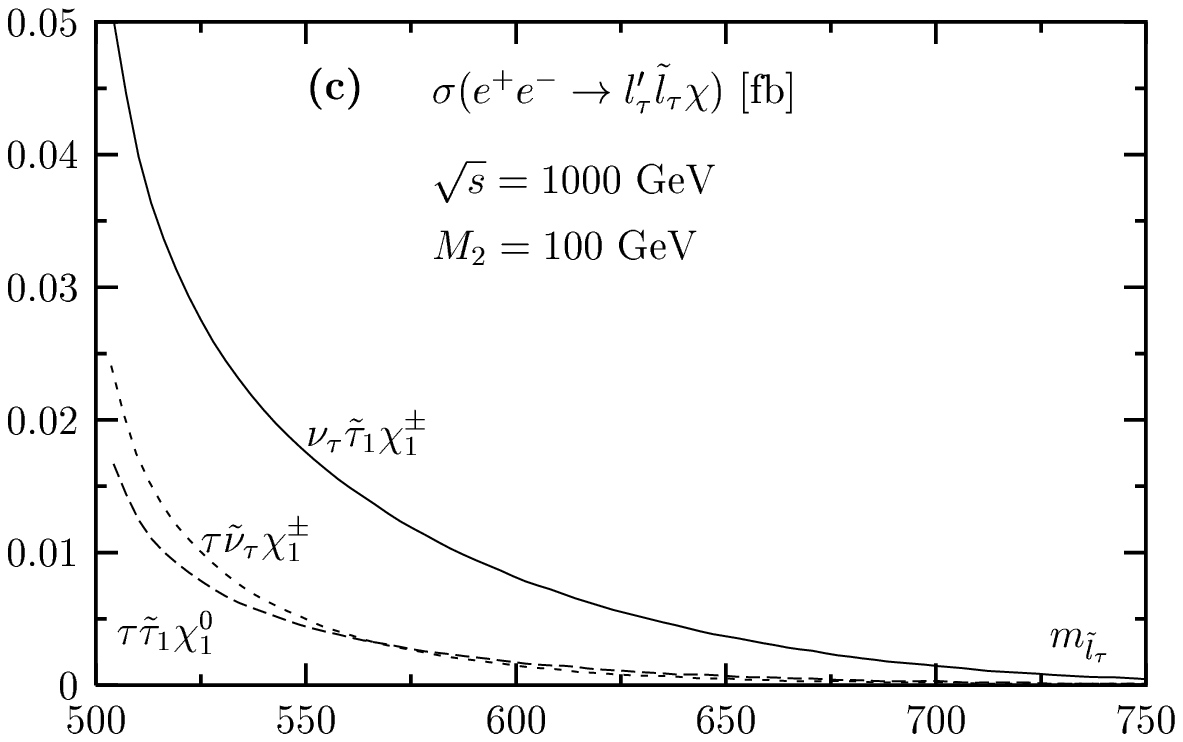,width=17cm}}
\vskip-14cm
\caption{Associated production cross sections for sleptons and gauginos 
in $\ee$ collisions for a c.m. energy $\sqrt{s}=1$ TeV as a function 
of the slepton masses. For the SUSY parameters, we have set $m_{\tilde{\ell}_L}
= m_{\tilde{\ell}_R}, M_2=100$ GeV, $\mu=500$ GeV and $\tb=30$.}  
\end{center}
\end{figure}

The cross sections for the same associated production processes but for a
center of mass energy $\sqrt{s}= 1$ TeV are shown in Fig.~4 as a function of
the slepton masses for the same set of inputs as in Fig.~3. The production
rates are smaller than in the previous case, a consequence of the fact that the
dominant contributing channels are the $s$--channel diagrams, which scale like
$1/s$ at high--energy. This drop should however be partly compensated by the
increase of the luminosity with energy [as expected for the TESLA machine for
instance], allowing for a reasonable number of events for slepton masses 
a few tens of GeV above the kinematical threshold. 

Fig.~5 shows the associated production cross sections in the smuon sector in
two particular cases for a c.m. energy $\sqrt{s}=500$ GeV. In Fig.5a, the
cross sections for $\tilde{\mu}_L$ and $\tilde{\nu}_\mu$ production in
association with the heavier neutral gaugino $\chi_2^0$ are displayed [again,
$\tilde{\mu}_R$ does not couple to $\chi_2^0$ in this case].  Approximately,
they are the same and equal to the ones with LSP final states close to the
kinematical thresholds, despite the fact that $m_{\chi_2^0} \simeq 2
m_{\chi_1^0}$. In Fig.~5b, we show the cross section for a fixed slepton mass
of $m_{\tilde{\ell}}=275$ GeV and a varying $M_2$ parameter. While the
variations of $\sigma(\ee \to \mu \tilde{\mu}_L \chi_1^0)$ and $\sigma(\ee \to
\mu \tilde{\mu}_L \chi_2^0)$ are mild, the cross section $\sigma(\ee \to \nu_\mu
\tilde{\mu}_L \chi_1^+)$ drops quickly with increasing $M_2$. Therefore, the
cross sections for the associated production of smuons with heavier chargino
states will be in general smaller than what was shown in Figs.~3 and 4.  

\begin{figure}[htbp]
\begin{center}
\vskip-4cm
\hskip-1cm\centerline{\epsfig{file=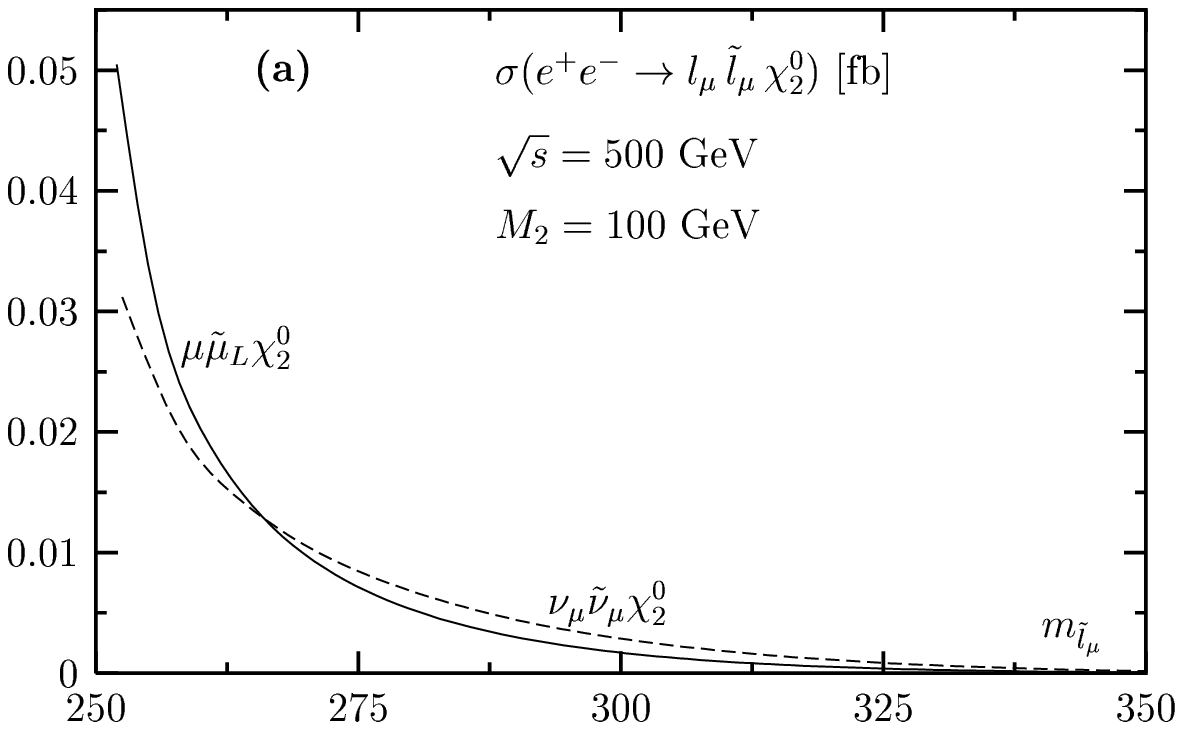,width=17cm}}
\vskip-17.5cm
\hskip-1cm\centerline{\epsfig{file=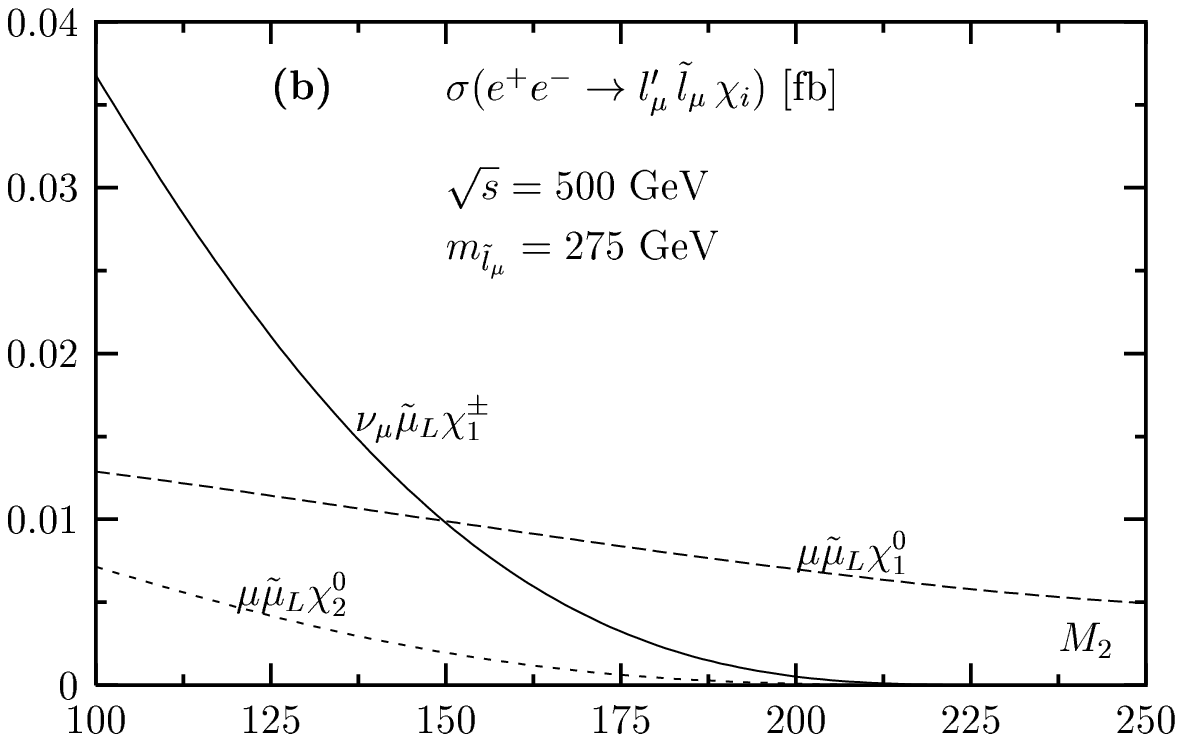,width=17cm}}
\vskip-14.5cm
\end{center}
\caption{Associated production cross sections for sleptons and gauginos
in $\ee$ collisions for a c.m. energy $\sqrt{s}=500$ GeV as a function of 
the slepton masses with $\chi=\chi_2^0$ (a) and as a function of the gaugino 
mass parameter $M_2$ with $\chi=\chi^0_{1,2}$ (b).}
\end{figure}

Finally, Fig.~6 shows the production cross sections for the case where the
lightest neutralino is higgsino--like in Fig.~6a [with $\mu=100$ GeV and 
$M_2=500$ GeV] or a mixture of higgsino and gaugino states in Fig.~6b 
[with $M_2=\mu=150$ GeV leading to $\chi_1^+$ masses $m_{\chi_2^\pm} \sim 
2 m_{\chi_1^\pm} \sim 200$ GeV]. As mentioned earlier, the cross sections are 
extremely small in the higgsino case and the situation should be the same for 
the production of smuons with $\chi_2^0$ and $\chi_1^\pm$,  and also for the 
production of $\tilde{\tau}$'s with higgsinos for not too large values of $\mu$
and $\tb$. However, for the mixed case, the cross sections are only slightly 
smaller than in the gaugino case. This is due to to the fact that all 
neutralino and chargino states [which share the gaugino couplings] have to be 
taken into account, leading to an overall contribution which is not very far 
from the one for the pure gaugino case. \s

\begin{figure}[htbp]
\begin{center}
\vskip-4cm
\hskip-1cm\centerline{\epsfig{file=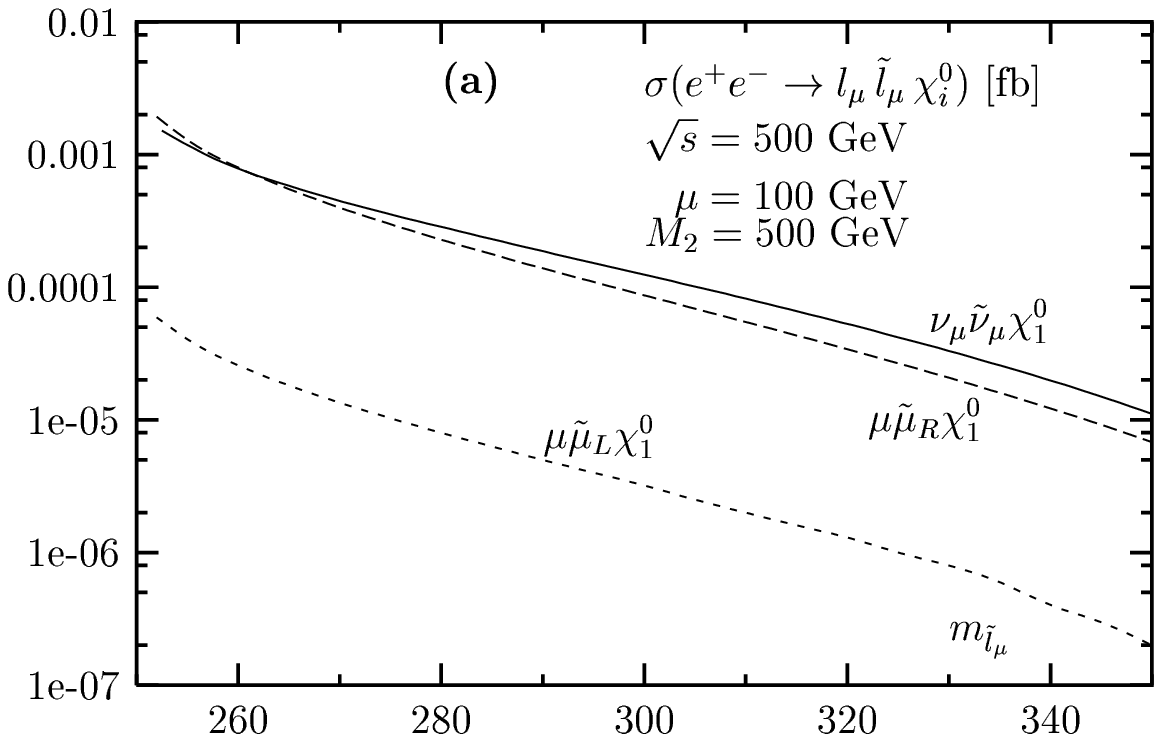,width=17cm}}
\vskip-17.5cm
\hskip-1cm\centerline{\epsfig{file=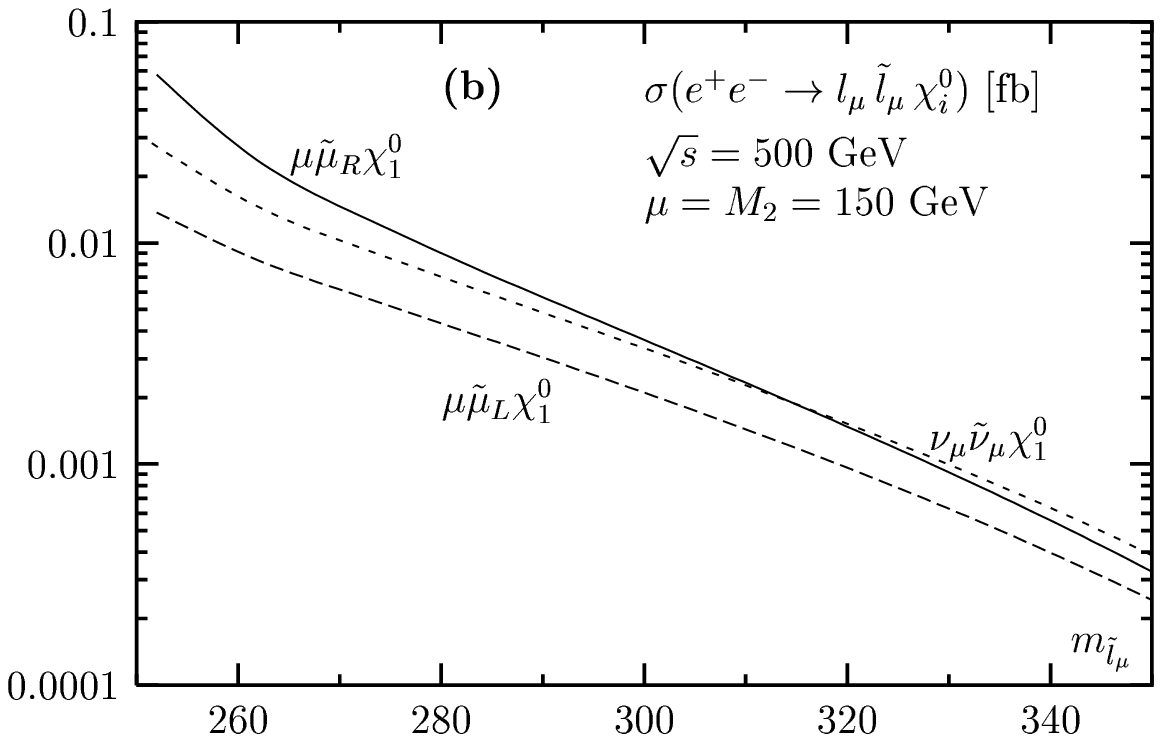,width=17cm}}
\vskip-14.5cm
\end{center}
\caption{Associated production cross sections for sleptons and gauginos
in $\ee$ collisions for a c.m. energy $\sqrt{s}=500$ GeV as a function of 
the slepton masses for the higgsino case (a) and the mixed gaugino--higgsino
case (b).} 
\end{figure}

Let us now turn to the case of $\tilde{t}$ and $\tilde{b}$ squarks. The cross
sections for the associated production of these squarks with the lightest
neutralino and chargino are shown in Fig.~7 for center of mass energies of 500
GeV (a), 1 TeV (b) and 3 TeV (c). In the 500 GeV case where phase space plays
an important role due to the heaviness of the top quark, the largest cross
section is obtained in the case of $\ee \to b \tilde{b}_1 \chi_1^0$ but it 
barely reaches the level of 0.03 fb for $m_{\tilde{b}_1} \sim 260$ GeV, leading
to not much more than 10 events for a luminosity of $\int {\cal L} \sim 500$
fb$^{-1}$. It is followed by the cross section for $\ee \to b \tilde{t}_1
\chi_1^\pm$ which is approximately a factor of two smaller. The cross section
for  $\ee \to t \tilde{t}_1 \chi_1^0$ is one order of magnitude smaller because
of the strong suppression by the phase space [since for $m_{\tilde{t}_1} \gsim
250$ GeV, $m_t+ m_{\tilde{t}_1}+ m_{\chi_1^0} \gsim  475$ GeV, i.e. close to
$\sqrt{s}=500$ GeV] while the process $\ee \to t \tilde{b}_1 \chi_1^\pm$ is not
kinematically accessible at this energy.  \s

At higher energies, Figs.~7b and 7c, the largest cross section is $\sigma( \ee
\to b \tilde{t}_1 \chi_1^\pm)$, which at $\sqrt{s}=1$ TeV can reach the level
of 0.1 fb for a stop mass of a few tens of GeV above the kinematical threshold,
followed respectively, by the cross sections $\sigma (\ee \to t \tilde{t}_1
\chi_1^0)$, $\sigma (\ee \to t \tilde{b}_1 \chi_1^\pm)$ and $\sigma (\ee \to b
\tilde{b}_1 \chi_1^0)$ which, up to a factor of two, have the same size. This
hierarchy of cross sections near the threshold follows from the dominance of 
final states with stop squarks since the main contribution comes from the
$s$--channel production of stops with photon--exchange [because of the larger
$Q_{\tilde{t}}$] and by the fact that charged current processes are more
favored. At even higher energies, Fig.~7c, the trend is similar to the previous
case, but the cross sections are smaller since the contributions of the dominant
$s$--channels scale like $1/s$.  \s

Finally, the cross sections for the associated production of third generation
squarks with gluinos are shown in Fig.~8. Here, there are two kinematical
situations which are relevant for a (genuine) three--body particle final state:
$(i)$ if the gluino is heavier than squarks and in this case, it is the only
way to produce gluinos in $\ee$ collisions [except for the loop induced
production of gluino pairs \cite{gluino} for which the cross sections are too
small] and $(ii)$ if the gluinos are lighter than squarks, with the latter not
being kinematically accessible in pairs in $\ee$ collisions [otherwise, one can
produce squarks which then decay into a quark and a gluino pair, with a
branching ratio that is dominant since it is a strong interaction process]. 
The cross sections for gluino production in association of $\tilde{b}_1$ and
$\tilde{t}_1$ squarks are shown for c.m. energies of 1 TeV (Figs.~8a,8c) and 3
TeV (Figs.~8b,8d) in these two kinematical regimes. In the first situation
[gluino heavier than squarks], the production cross sections can reach the
level of a fraction of a femtobarn leading to a few tens of events for the
expected luminosities, $\int {\cal L} \sim 500$ fb$^{-1}$. In the second
situation [squarks heavier than gluinos], the cross sections are rather tiny
[below 0.01 fb] and higher luminosities are needed to detect these final
states.  

\begin{figure}[htbp]
\begin{center}
\vskip-4cm
\hskip-1cm\centerline{\epsfig{file=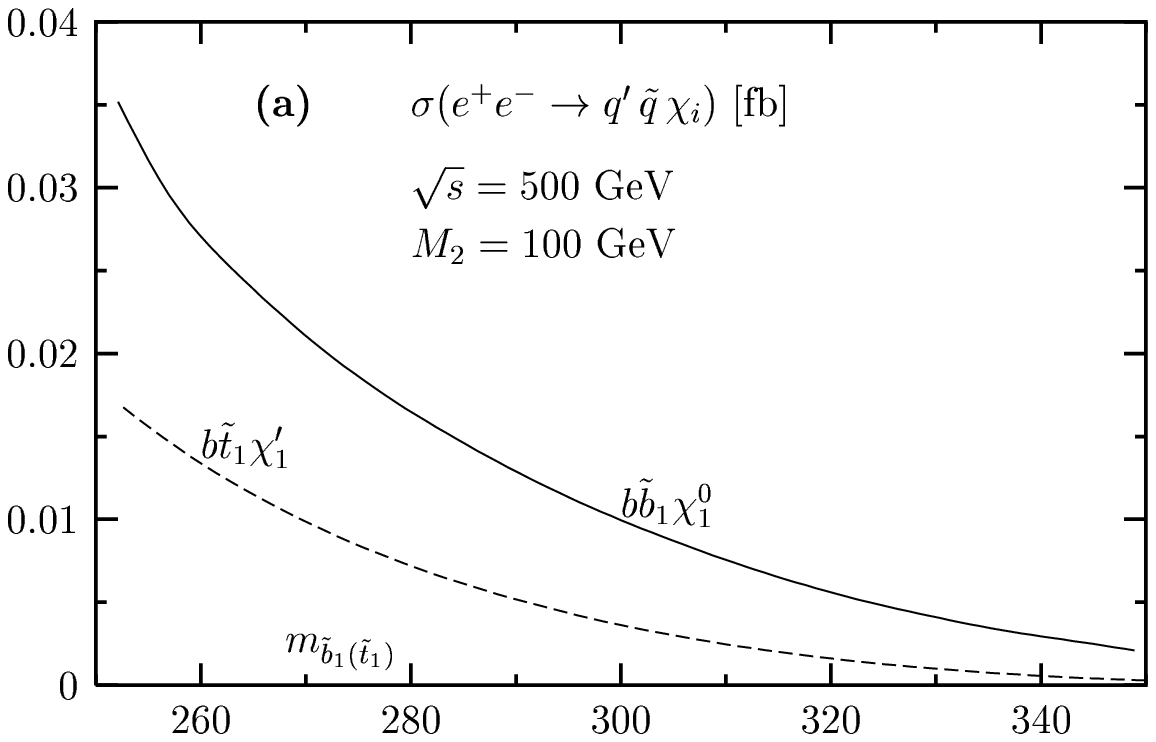,width=17cm}}
\vskip-17.5cm
\hskip-1cm\centerline{\epsfig{file=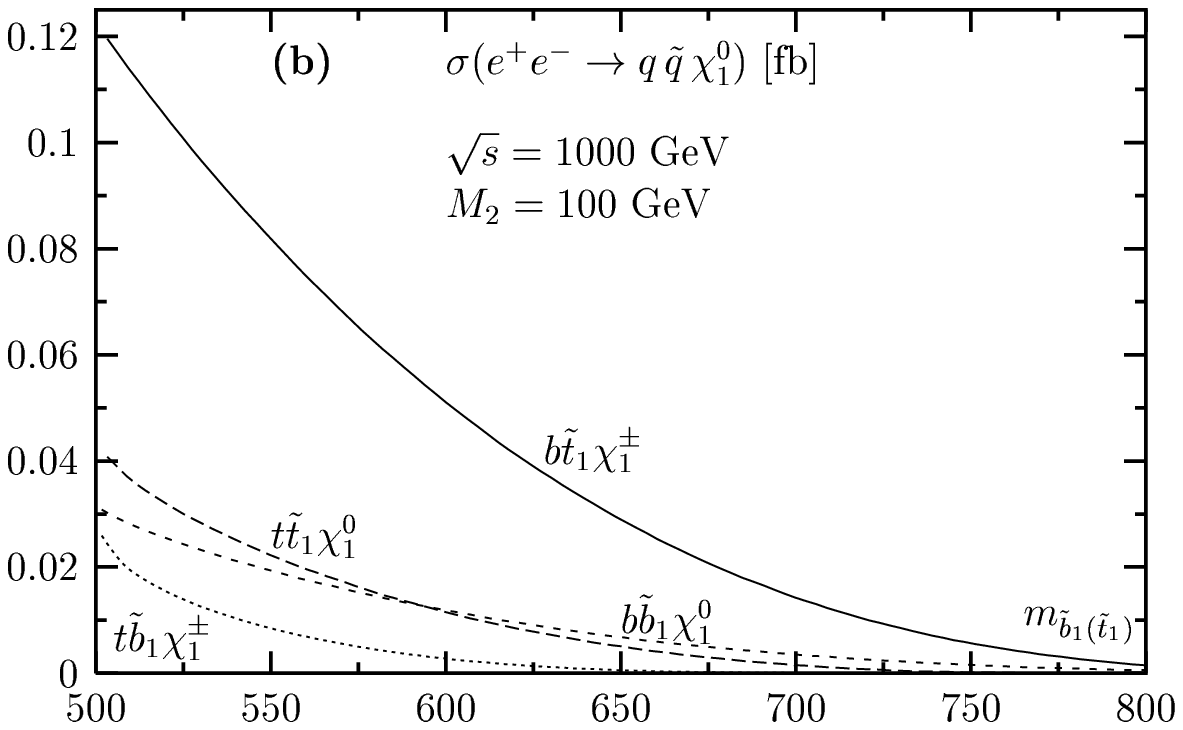,width=17cm}}
\vskip-17.5cm
\hskip-1cm\centerline{\epsfig{file=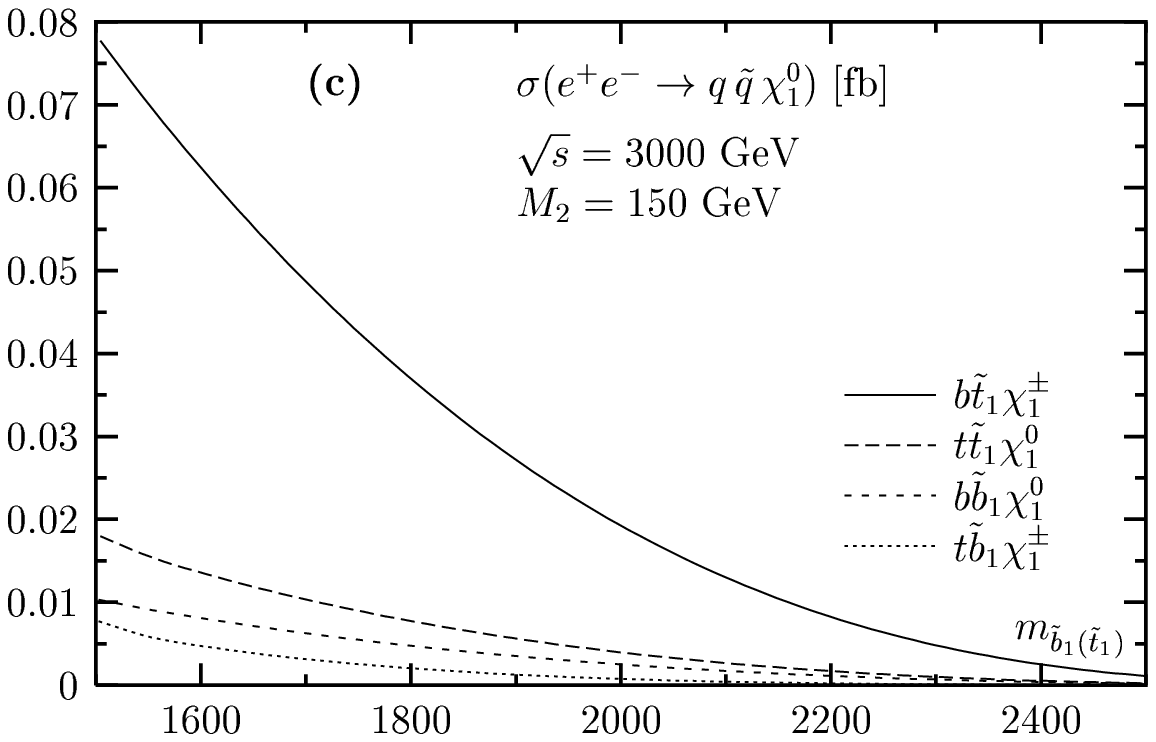,width=17cm}}
\vskip-14cm
\caption{Associated production cross sections for squarks and electroweak 
gauginos in $\ee$ collisions for a c.m. energies of $\sqrt{s}=500$ GeV (a),
1 TeV (b) and 3 TeV (c) as a function of the squark masses. For the SUSY 
parameters, we have set $m_{\tilde{q}_L} = m_{\tilde{q}_R}, M_2=100$ GeV 
(150 GeV in c), $\mu=500$ GeV and $\tb=30$.}  
\end{center}
\end{figure}

\begin{figure}[htbp]
\begin{center}
\vskip-6.cm
\mbox{\hskip-3cm\centerline{\epsfig{file=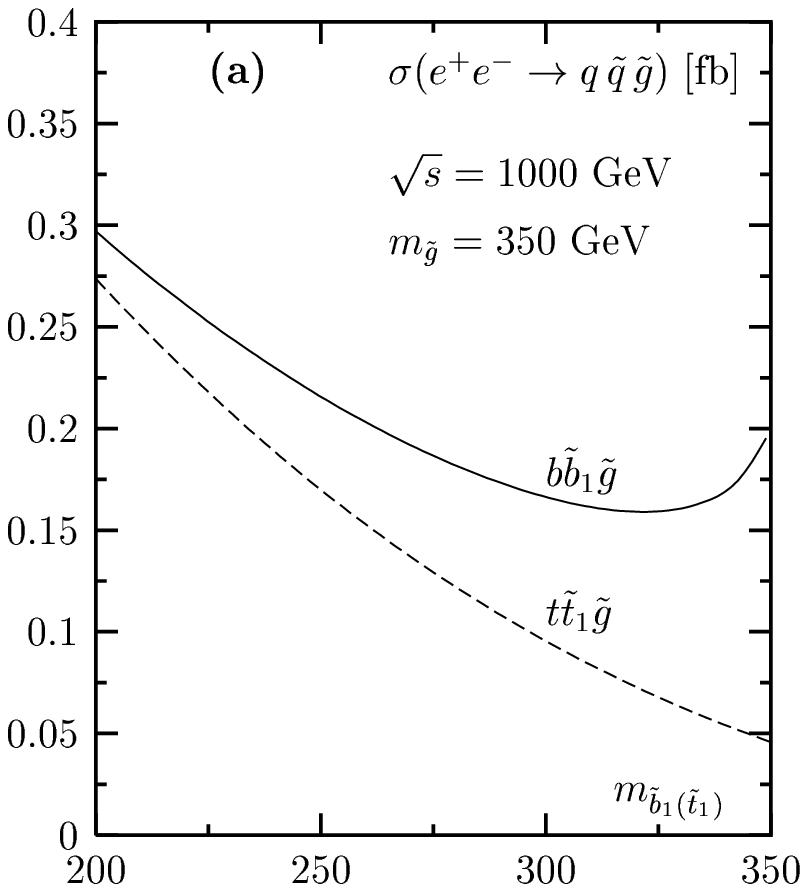,width=18.5cm}}
\hskip-8.5cm\centerline{\epsfig{file=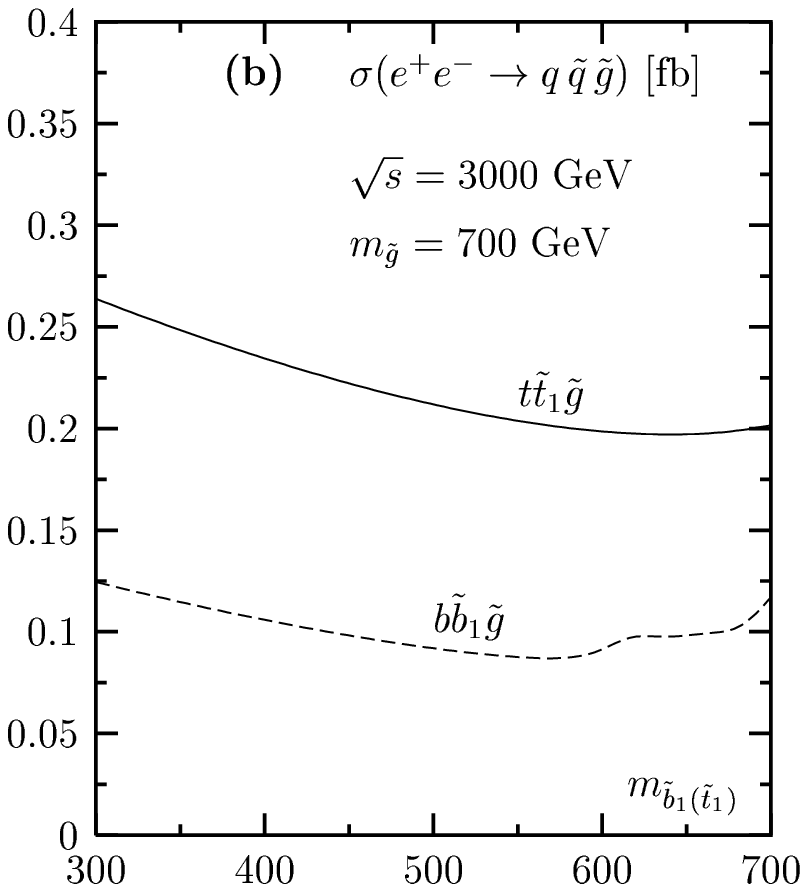,width=18.5cm}} }
\vskip-17.5cm
\mbox{\hskip-3cm\centerline{\epsfig{file=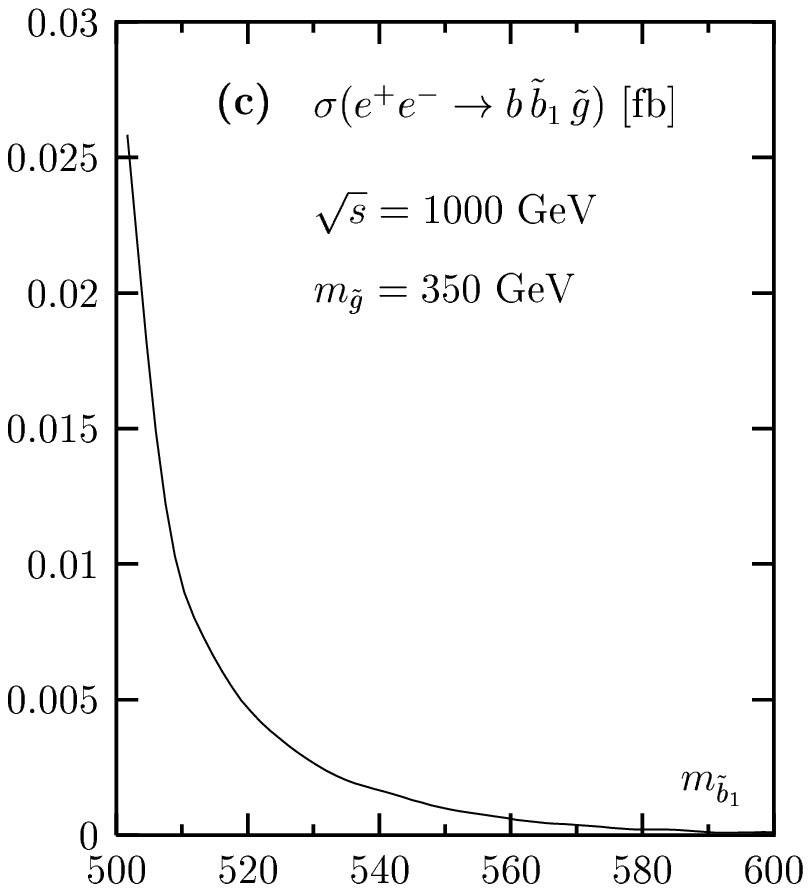,width=18.5cm}}
\hskip-8.5cm\centerline{\epsfig{file=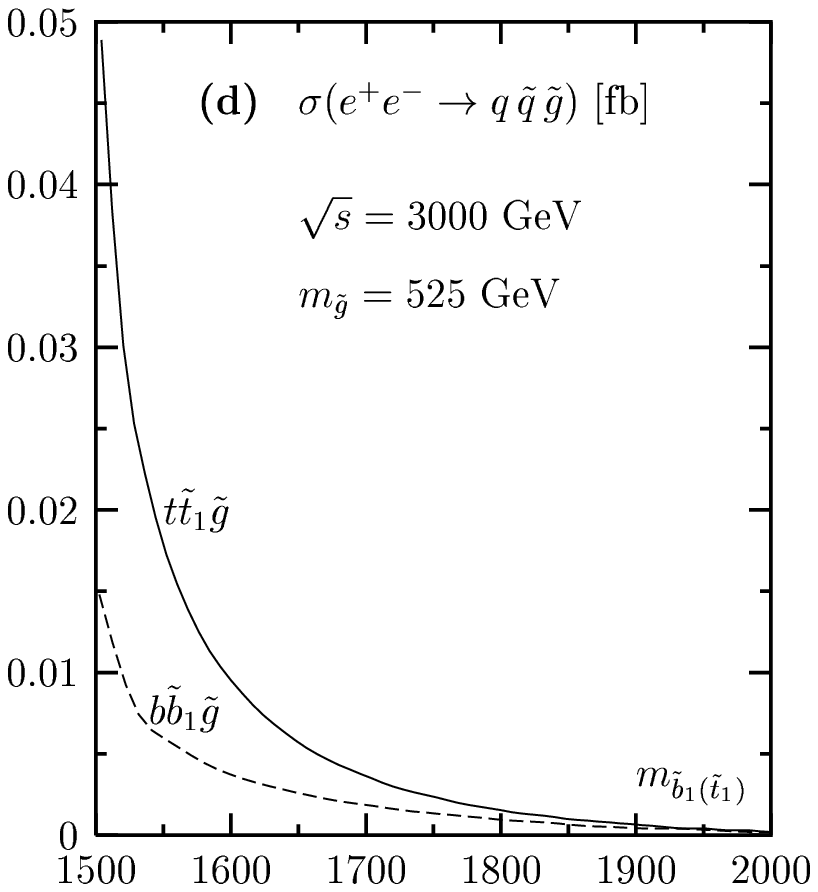,width=18.5cm}} }
\vskip-13.99cm
\end{center}
\caption{Associated production cross sections for $\tilde{b}_1$ and
$\tilde{t}_1$ squarks and gluinos in $\ee$ collisions for a c.m. energy
$\sqrt{s}=1$ TeV (a,c) and 3 TeV (b,d) as a function of the squark masses.  For
the SUSY parameters, we have set $M_2=350$ GeV (525 GeV) in a-c (b-d),
$\mu=500$ GeV and $\tb=30$. The increase of the cross section in (a) for
sbottoms for $m_{\tilde{b}_1} \sim m_{\tilde{g}}$ is due to the small
virtuality of $\tilde{b}_1$ in this case.  The kink in (b) is due to the
heavier $\tilde{b}_2$ (exchanged in the $s$--channel digram with a $Z$) becoming
close in mass to the gluino. For stops, this is not apparent in the curves 
because of the large value of $m_t$.} 
\end{figure}

\subsection*{4. The cross sections in various approximations}

As mentioned in section 2.3 and 3.1, the bulk of the production cross sections
for the associated production of sfermions, fermions and electroweak gauginos
is due to the contributions of the $s$--channel diagrams, Figs.~1a--1c. In
particular, near the kinematical threshold for sfermion pair production,
$\sqrt{s} \simeq 2 m_{\tilde{f}}$, the contribution of digram (1a) with an
almost resonant sfermion is dominant.  However, this diagram is not gauge
invariant by itself, although in a covariant gauge, the gauge parameter 
dependence drops out. 
Furthermore, very close to the kinematical threshold, the zero--width
approximation is not valid anymore and one has to include the total
decay width of the sfermion in the Breit--Wigner form of the resonant
propagator. This is performed by substituting in the sfermion propagator, the
squared mass $m_{\tilde{f}}^2$ by $m_{\tilde{f}}^2-i m_{\tilde{f}}
\Gamma_{\tilde{f}}$.  In fact, near this kinematical threshold, the three--body
approximation is not accurate anymore and for a more reliable result one has to
consider the four--body process with the production of two off--shell sfermions
which then split [or decay if one is above threshold] into fermions and
gauginos.  Here again, the sole contribution of this double resonant diagram
[which, in general, is the only one taken into account in experimental
analyses, see Ref.~\cite{Uli} for instance] is not gauge invariant and one
should take into account additional diagrams with single resonant states [i.e.
with only one intermediate sfermion state or with neutralino or chargino
states] \cite{Peter}. \s 

In this section, we will therefore discuss in more details the magnitude of
the production cross sections in the various approximations taking as example
the associated production of the right--handed smuon, a muon and the LSP,
$\ee \to \mu^+ \tilde{\mu}_R \chi_1^0$ and the charge conjugate final state
$\mu^- \tilde{\mu}_R^* \chi_1^0$. The total decay width of the 
right--handed smuon, assuming that the LSP is a pure bino [which implies that 
$\tilde{\mu}_R$ does not couple not only to the heavier charginos and 
neutralinos which are higgsino--like but also to the gauginos $\chi_1^\pm$ and 
$\chi_2^0$],  is identical to the partial decay width into a muon 
and the LSP which is given by [the couplings given in section 2.2 have been
simplified]
\beq
\Gamma_{\tilde{\mu_R}} = \frac{1}{2} \alpha \bigg( N'_{11}- \frac{s_W}{c_W}
N'_{12} \bigg)^2 \, m_{\tilde{\mu}_R} \, \bigg(1- \frac{ m_{\chi_1^0}^2}
{m_{\tilde{\mu}_R}^2} \bigg)^2  
\eeq
For a 250 GeV $\tilde{\mu}_R$, assuming as usual $M_2=100$ GeV with $\tb=30$ 
and $\mu=500$ GeV, the total decay width is $\Gamma_{\tilde{\mu}_R}\simeq 1.15$ 
GeV. \s

\begin{figure}[htbp]
\begin{center}
\vskip-5.5cm
\hskip-1cm\centerline{\epsfig{file=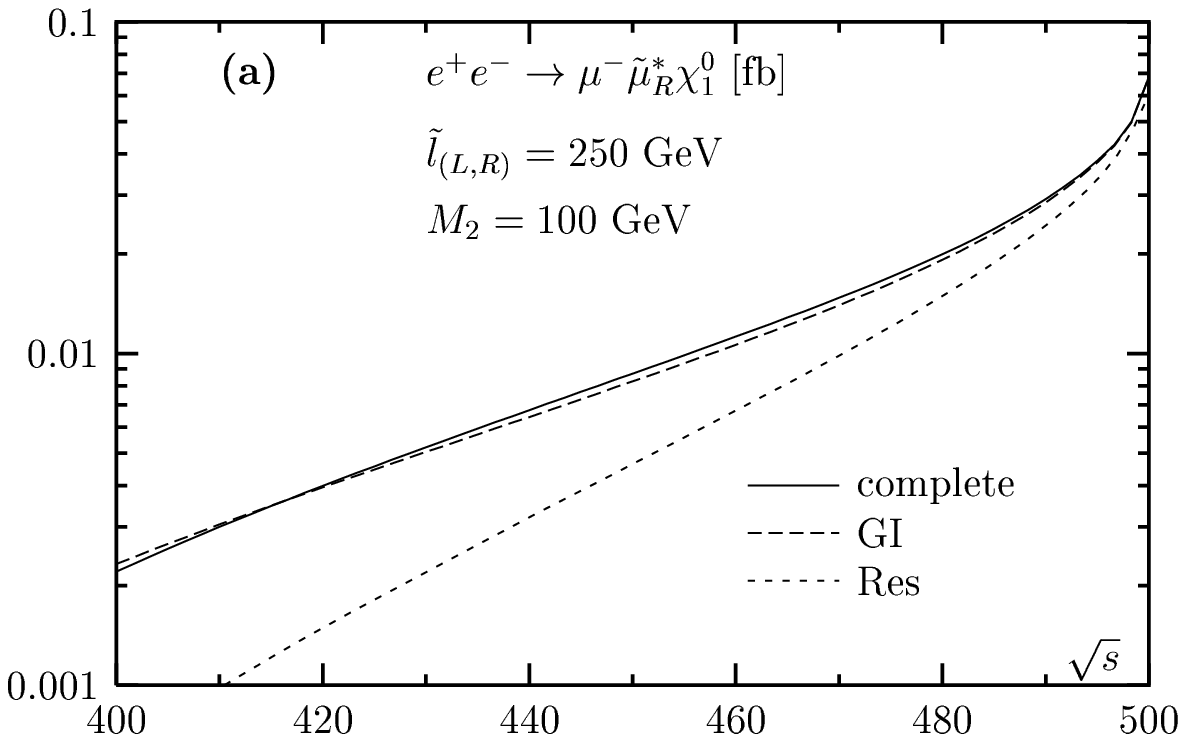,width=24cm}}
\vskip-23.3cm
\mbox{\hskip-3cm\centerline{\epsfig{file=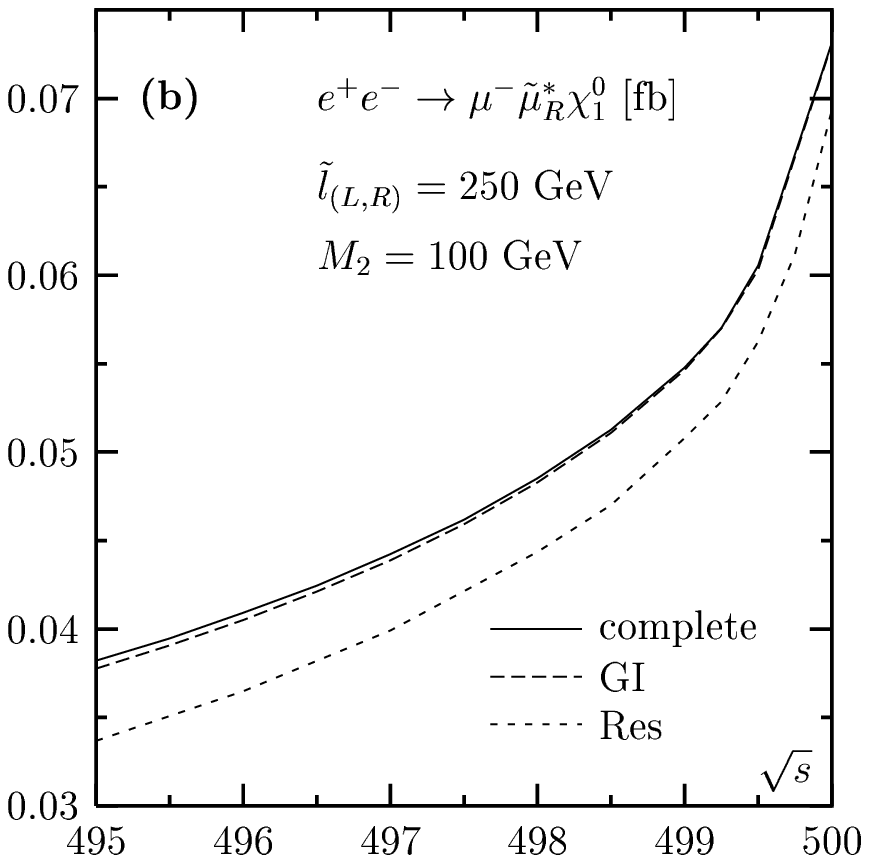,width=18.5cm}}
\hskip-8.5cm\centerline{\epsfig{file=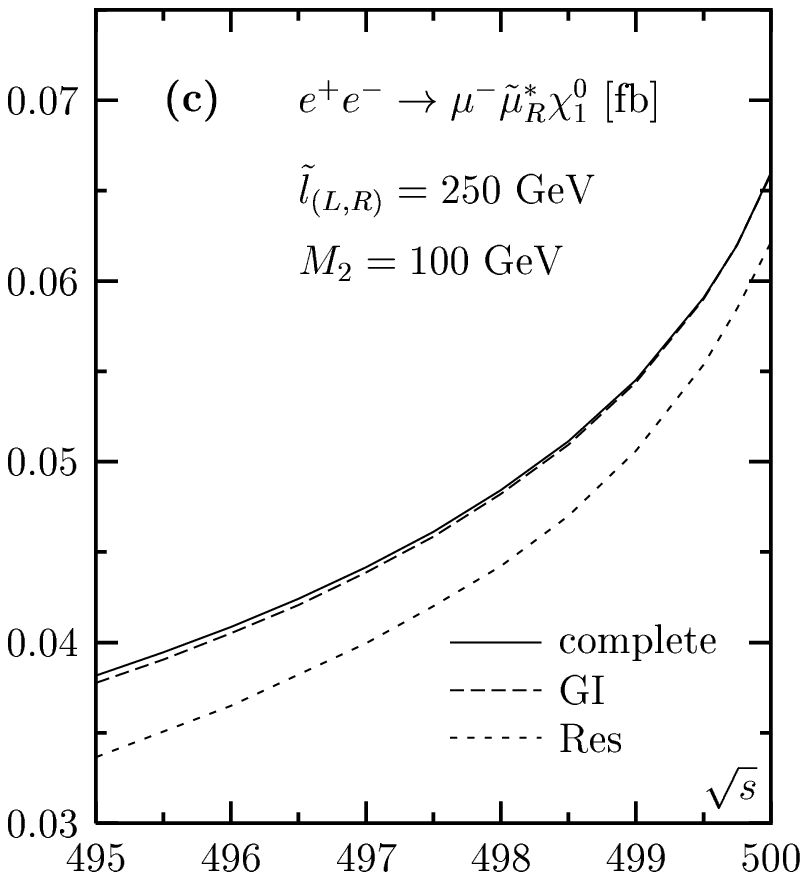,width=18.5cm}} }
\vskip-14cm
\end{center}
\caption{The total cross section for the associated production process $\ee \to
\mu^- \tilde{\mu}_R^* \chi_1^0$ as a function of the center mass energy in the
exact case (solid lines) and in the approximations of the gauge invariant set
of diagrams (dashed lines) and the resonant smuon diagram (dotted lines). In
panels (a) and (c) the total smuon width is included, while in panel (b) it is
set to zero.} \end{figure}

In Fig.~9a, we show the cross section for this process as a function of the
c.m.~energy [varied from 400 to 500 GeV] for $m_{\tilde{\ell}_{L,R}}=250$ GeV
and the SUSY parameters given above, retaining $\Gamma_{\tilde{\mu}_R}$.  While
the short--dashed curve shows the contribution of the diagram (1a), the
long--dashed curve shows the contribution of the three $s$--channel diagrams
(1a), (1b) and (1c), and the full line shows the total contribution including
all diagrams. \s

As one can see, it is a very good approximation [in the present case] to
include only the gauge invariant set of contributions (diagrams 1a+1b+1c),
since the full line and the long--dashed line are almost overlapping in the
entire range of $\sqrt{s}$ values shown in the figure. Far below the
kinematical threshold, the contribution of diagram (1a) is rather small, but
closer to the threshold where the exchanged smuon is almost on mass--shell, it
becomes the dominant one. This is shown in a more explicit way in Fig.~9b,
where we zoom on the threshold region.  A few GeV near $\sqrt{s}=2
m_{\tilde{\mu}_R}$, the contribution of the almost--resonant smuon exchange
diagram represents 95\% of the total cross section. In Fig.~9c, we display the
effect including the total width of the smuon in the resonant diagram (1a).  As
can be seen, the production cross section is slightly lower [by $\sim 5\%$]
compared to the zero--width case, Fig.~9b.  Therefore, close to the kinematical
threshold, one has to include all the $s$--channel diagrams as well as the
finite width of the smuon to have a rather accurate estimate of total cross
section. \s

Let us now turn to the discussion of the four--body production process, $\ee
\to \mu^+ \chi_1^0 \mu^- \chi_1^0$. Taking into account only the gauge
invariant set of the double--resonant diagram, $\ee \to \tilde{\mu}_R^*
\tilde{\mu}_R^* \to \mu^+ \chi_1^0 \mu^- \chi_1^0$, plus the single resonant
ones with either smuons or gauginos in the intermediate states, 10 Feynman
diagrams [the generic form of which can be obtained from the set of diagrams
(1a), (1b) and (1c) of Fig.~1, where the smuon splits into a muon and a
neutralino] are to be considered. We have calculated the total production cross
section, including the finite widths of the smuons, in this approximation using
the package {\tt CompHEP} \cite{CompHep}. This should be a good
approximation\footnote{The full calculation of the four--body final state
process involves 84 Feynman diagrams. Even if one neglects the diagrams
involving the couplings $Z$--$\chi^0$--$\chi^0$, $\ell$--$\tilde{\ell}_R
$--$\chi_2^0$ [which vanish in the gaugino LSP limit in a covariant gauge] and
$Z$--$\tilde{\mu}_1$--$\tilde{\mu}_2$ [which is zero in the no--mixing case],
44 Feynman diagrams remain. Even with {\tt CompHEP}, the calculation in this
simpler case was consuming an enormous amount of time and we did not perform
it. This is partly due to the huge volume of the kernel matrix element squared
and partly to the Monte--Carlo integration on the 4--body final state which
calls for a large number of points and iterations to have a better convergence
of the integral.} corresponding to what occurs in the three--body production
process.  We have compared our results with the ones obtained in
Ref.~\cite{Peter} and found perfect agreement. \s

The four--body cross section, as a function of the c.m. energy, is shown by the
solid line of Fig.~10 for the set of inputs chosen previously. It is compared
with the cross section for the three--body process below and above
threshold\footnote{Note that for the three--body process, we include for the
below threshold case, as usual, both the cross section for $\tilde{\mu}_R$ and
for the conjugate state $\tilde{\mu}_R^*$. Above the threshold, we have to
divide the cross section by a factor $\frac{1}{2}$ to take into account the
fact that, after the decay of the (on--shell) smuons, we have two identical neutralinos in
the final state.} [long--dashed lines]  and with the cross section of the
two--body $\ee \to \tilde{\mu}_R^+ \tilde{\mu}_R^-$ production above the
threshold with the subsequent decays of the smuons into $\mu \chi_1^0$ final
states [short--dashed lines]. The on--shell smuons are assumed to decay 100\%
of the time into $\mu \chi_1^0$ final states. As can be seen, below the
kinematical threshold the four--body and the three--body cross section are
almost overlapping, which means that it is really a very good approximation to
consider in this regime the much simpler three--body production case. 
Sufficiently above the kinematical threshold, the three--curves also overlap as
it should be. A few GeV above the threshold, the three--body and two--body
approximations fail badly, and one has to consider the full process with two
virtual smuons including their total decay widths in the
propagators\footnote{Of course, one has to consider the full contribution of
the gauge invariant set discussed above. For $\sqrt{s}=485$ GeV, this gives a
total cross section of 0.046 fb for the set of inputs of Fig.~10, while the
inclusion of only the doubly resonant diagram gives a cross section of only
0.037 fb.}. \s

In fact, for a better description of the cross section around the kinematical
threshold, which is needed to meet the experimental possibility of measuring
the produced smuon mass with a scan near threshold with a precision better than
100 MeV \cite{TESLA}, one has not only to use the full four--body production
process, $\ee \to \mu^+ \mu^- \chi_1^0 \chi_1^0$, including the smuon finite
widths, but also to take into account some refinements such as Coulomb
re-scattering effects, initial state radiation from the emission of collinear
and soft photons and beamstrahlung effects, as discussed in Ref.~\cite{Peter}.  

\begin{figure}[htbp]
\begin{center}
\vskip-6cm
\hskip-1cm\centerline{\epsfig{file=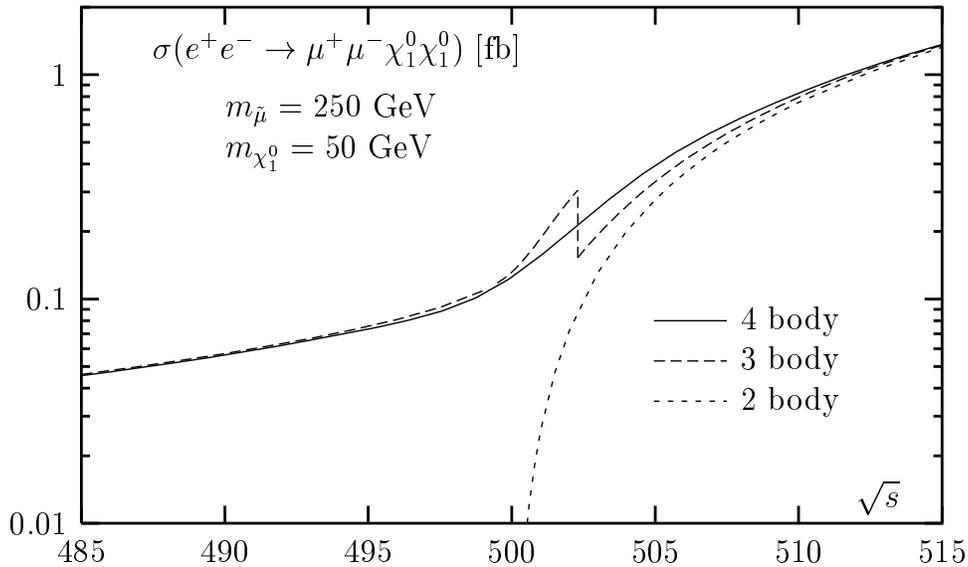,width=22cm}}
\vskip-18.5cm
\end{center}
\caption{Cross section for the two--, three-- and four--body processes leading
to the final state $\mu^+ \mu^- \chi_1^0 \chi_1^0$ with intermediate
right--handed smuons, around a c.m.~energy of 500 GeV. For the SUSY parameters,
we have set $M_2 \simeq 2M_1 = 100$ GeV, $\mu=500$ GeV and $\tb=30$. For the 
three--body curve, the kink at $\sqrt{s}\simeq 2(m_{\tilde{f}}+\Gamma_{\tilde{f}
})$ is due to the ``brutal" inclusion of the factor $\frac{1}{2}$ to avoid 
double counting above threshold; the approximation slightly above threshold is 
thus rather bad (see text). }
\end{figure}

\subsection*{5. Conclusion}

In this paper, we have analyzed the production at future $\ee$ colliders of
second and third generation sleptons and third generation squarks, in
association with their partner leptons and quarks and charginos or neutralinos
[and gluinos in the case of squarks], $\ee \to f \tilde{f} \chi$, in the
framework of the MSSM. Analytical expressions for the differential cross
sections have been given in the approximation where only the $s$--channel
contributing diagrams are taken into account, an approximation which has been
shown, a posteriori, to be very good.  Taking the example of right--handed
smuon production, we have shown that consideration of only the three--body
processes is a very good approximation below the kinematical two--sfermion
production threshold, but that close to these threshold the full four--body
process $\ee \to f \bar{f} \chi \chi$, including the finite total decay width
of the sfermions, have to be considered \cite{Peter}.  \s

We have shown that some of these three--body processes can have sizeable enough
production cross sections to allow for the possibility of discovering sfermions
with masses a few tens of GeV above the kinematical threshold for sfermion pair
production $\sqrt{s}=2m_{\tilde{f}}$ in favorable regions of the SUSY parameter
space. This is due to the very high luminosities expected at future linear $\ee$
colliders, $\int {\cal L} \sim 500$ fb$^{-1}$, i.e. three orders of magnitude 
larger than in the case of LEP2 where the experimental limits on the sfermions 
masses do not exceed the beam energy. Associated production of squarks and
gluinos offers, for some range of sparticle masses, the unique direct access 
to the gluinos in $\ee$ collisions, if they are (slightly) heavier than 
squarks. \s

The final states discussed in this paper should be clear enough to be easily
detected in the clean environment of $\ee$ colliders. However, for a more
precise description, detailed analyses [which are beyond the scope of the 
present paper] which take into account Standard Model and SUSY backgrounds [as 
those performed in Ref.~\cite{Peter} in the case of smuon production near 
threshold] as well as higher order effects and detection efficiencies, have to 
be performed. 

\bigskip

\nn {\bf Acknowledgments}: \s

\nn We thank Ayres Freitas and Peter Zerwas for clarifying discussions, Edward
Boos and Andrei Semenov for their help in using {\tt CompHEP} and Yann Mambrini
and Margarete M\"uhlleitner for discussions on Majorana neutralinos. A.~Datta
is supported by a MNERT fellowship. This work is supported in part by the
Euro--GDR Supersym\'etrie and by the European Union under contract
HPRN-CT-2000-00149.  

\newpage
\subsection*{Appendix: the differential cross section}
\setcounter{equation}{0}
\renewcommand{\theequation}{A.\arabic{equation}}

In this appendix, we present the analytical expression of the differential 
cross section for the associated production of sfermions with charginos or 
neutralinos
\beq
e^+(p_1) \; e^-(p_2) \to \tilde{f}_h (p_3) \; \chi_i(p_4) \; 
\bar{f}^\prime(p_5) 
\eeq
where $h=L,R$ is the handedness of the produced sfermion [later on, we will
use $h'=R,L$ which is the other possible helicity]. To simplify the
expressions, we will work in the approximation in which the accompanying fermion
in the final state is massless, leading to zero--mixing in the sfermion
sector [these expressions are therefore not accurate in the case of associated
production with final state top quarks]. Furthermore, we will take into account
only the contributions of the gauge invariant set of diagrams Fig~1a, 1b and 1c
with $s$--channel exchange of photons and $Z$ bosons which, as shown in the
numerical analysis in section 4, is a very good approximation to the full 
cross section.  \s
 
The complete spin-averaged matrix element squared for vanishing final state 
fermion mass and zero--mixing in the scalar sector, is given by 
\begin{equation}
|{\cal M}|^2 = 4 N_c \frac{(4\pi \alpha)^3}{\stwn2} \; \sum_{i \geq j=a,b,c} 
T_{ij}
\end{equation}
where $N_c$ is the color factor and $\alpha$ is the fine structure constant.
The $T_{ij}$ are the amplitudes squared with $i=j=a,b,c$ of the contributions 
of diagrams $1a,1b$ or $1c$ where the $\gamma$ and $Z$ boson amplitudes have 
been added, and the interference terms are obtained for $i \neq j=a,b,c$; here 
the effects of permutations of suffixes are already included in the original 
terms and hence such permutations are to be left out. \s

The diagonal terms are given by: \s 

\beq
T_{aa} &=& \frac{\aih^2}{\{(q-\p3)^2-\msfh^2 \}^2} 
\left [ \frac{\qf^2}{s^2} + \frac{\azff^2}{\stwn4 \ctwn4} 
\frac{(\ae^2+\ve^2)}{(s-\mz^2)^2} - 2 \frac{\qf \azff \ve}{\stwn2 \ctwn2 s
(s-\mz^2)} \right ] K_1
\eeq
\beq
T_{bb} = \frac{\aih^2}{\{(q-\p5)^2-\mfer^2 \}^2}
&\Biggl[& \hskip -10pt
\left \{ \frac{\qf^2}{s^2} + \frac{(\ae^2+\ve^2) 
(\af-\vf)^2}{\stwn4 \ctwn4 (s-\mz^2)^2} + \frac{2 \qf (\af-\vf) 
\ve}{\stwn2 \ctwn2 s (s-\mz^2)} \right \} K_2 \non \\ 
&-& \frac{2 \ae (\af -\vf)}{\stwn2 \ctwn2 (s-\mz^2)} 
\left \{ \frac{\ve (\af-\vf)}{\stwn2 \ctwn2 (s-\mz^2)} + \frac{\qf}{s} \right \} K_2^\prime
\Biggr]
\eeq

\beq
T_{cc} &=& \frac{\qx^2}{s^2} \sum_j \frac{\ajh^2}{\left\{ (q-\p4)^2-\mgj^2 \right \}^2 }
\left \{ (\mgj^2-\msfh^2) K_3 + K_3^\prime + K_3^{\prime\prime} \right \} 
\non \\
&+& \hspace*{-4mm} \frac{(\ae^2+\ve^2)}{\stwn4 \ctwn4 (s-\mz^2)^2} 
\sum_k 
\frac{\akh^2 \left \{ (\mgk^2 {\ork}^2 - \msfh^2 {\olk}^2) K_3 + {\olk}^2 K_3^\prime +
\olk \ork K_3^{\prime\prime} \right \} }
{\left\{ (q-\p4)^2-\mgk^2 \right \}^2} \\
&+& \hspace*{-4mm} \frac{2 \ve \qx (-1)^{\qx}}{\stwn2 \ctwn2 s (s-\mz^2)}
\sum_{j,k}
\frac{\ajh \akh \left \{ (\mgk^2 \ork + \msfh^2 \olk) K_3 +  \olk 
K_3^{\prime} + \frac{1}{2}(\olk+\ork) K_3^{\prime\prime} \right \}}
{\left\{ (q-\p4)^2-\mgj^2 \right \} \left\{ (q-\p4)^2-\mgk^2 \right \} } 
\hspace*{-.5cm}
\non 
\eeq
while the non--diagonal terms are given by: 
\beq
T_{ab} &=& \frac{\aih^2}{\left\{ (q-\p3)^2-\msfh^2 \right \} \left\{ (q-\p5)^2-\mfer^2 \right \} }
\Biggl[-\frac{\qf^2}{s^2} K_{12} \non \\ 
&+&
\frac{1}{\stwn2 \ctwn2 (s-\mz^2)} \Biggl \{ \left \{ \frac{\azff (\af-\vf) (\ae^2+\ve^2)}
{\stwn2 \ctwn2 (s-\mz^2)} - \frac{\qf \ve}{s} (\af-\vf-\azff) \right \} 
K_{12}^{\prime\prime} \non \\
&-&
 \ae \left \{  \frac{2 \azff (\af-\vf) \ve}{\stwn2 \ctwn2 (s-\mz^2)} - \frac{\qf}{s}
(\af-\vf-\azff)  \right \} K_{12}^\prime
\Biggr\} \Biggr ] 
\eeq
\beq
T_{ac} &=& \frac{1}{(q-\p3)^2-\msfh^2} \sum_j \frac{\ajh^2}{(q-\p4)^2-\mgj^2}
\Biggl[ \frac{\qx}{s} \left \{ \frac{\qf}{s}-\frac{\azff \ve}{\stwn2 \ctwn2 
(s-\mz^2)} \right \} (K_{13} \non  \\ 
&& + \frac{(-1)^{\qx}}{\stwn2 \ctwn2 (s-\mz^2)} \left \{ 
\frac{\qf}{s} - 
\frac{\azff (\ae^2+\ve^2)}{\stwn2 \ctwn2 (s-\mz^2)} \right \} (\olj K_{13}
+\orj K_{13}^\prime) \Biggr]
\eeq
\beq
T_{bc} &=& \frac{1}{(q-\p5)^2-\mfer^2} \sum_j \frac{\ajh^2}{(q-\p4)^2-\mgj^2}
\Biggl[ \frac{\qx}{s} \left \{ \frac{\qf}{s} - \frac{(\af-\vf)\ve}{\stwn2 
\ctwn2 (s-\mz^2)} \right \} (K_{23}^a +K_{23}^b \non \\
&+& K_{23}^{c^\prime}) + \frac{(-1)^{\qx}}{\stwn2 \ctwn2 (s-\mz^2)} \left \{ 
\frac{(\af-\vf)(\ae^2+\ve^2)}
{\stwn2 \ctwn2 (s-\mz^2)} + \frac{\ve}{s} \right \} 
\left \{ \olj (K_{23}^a +K_{23}^{c^\prime}) + \orj K_{23}^b \right \} \non \\
&+& \frac{\ae (\af-\vf)}{\stwn2 \ctwn2 (s-\mz^2)} \left \{ (-1)^{\qx} \left \{ 
\frac{2\ve}{\stwn2 \ctwn2 (s-\mz^2)} + \frac{1}{s} \right \} 
(\orj K_{23}^{d^\prime}+\olj K_{23}^{d^{\prime\prime}}) -
\frac{\qx}{s} K_{23}^{d}
\right \} 
\Biggr] \non \\
\eeq
The factors $K$, which involve the dependence on the momenta, read: 
\beq
K_1 &=& \dotp45 (4 \dotp13 \dotp23 - \msfh^2 s) \non \\
K_2 &=& 4 \dotp15 \dotp25 \dotp45 + s \left \{ \dotp15(\dotp14-\dotp45)+
\dotp25(\dotp24-\dotp45)  \right \} \non \\
K_2^\prime &=& \dotp45(\dotp15-\dotp25)s \non \\
K_3 &=& \dotp14 \dotp25 + \dotp15 \dotp24 \non \\
K_3^\prime &=& (\dotp13 \dotp24 + \dotp14 \dotp23) \dotp35 \non \\
K_3^{\prime\prime} &=& \mgi \mgj s \; \dotp35 \non \\ \non \\
K_{12} &=& 4 (\dotp13 \dotp25 + \dotp15 \dotp23) \dotp45 
- s \, \left \{ (\dotp15+\dotp25) \dotp34 - (\dotp14+\dotp24) \dotp35 \right \} \non \\
K_{12}^\prime &=& s \, \left \{ (\dotp15-\dotp25) \dotp34 - (\dotp14-\dotp24) \dotp35 \right \}
\non \\
K_{12}^{\prime\prime} &=& K_{12} - 2 \dotp35 \dotp45 \, s \non \\ \non \\
K_{13} &=& 2 \left \{ (\dotp13 \dotp25 + \dotp15 \dotp23) \dotp34 - 
(\dotp13 \dotp24 + \dotp14 \dotp23) \dotp35 \right \} - (4 \dotp13 \dotp23 -\msfh^2 \, s) 
\dotp45 \non \\
K_{13}^\prime &=& \mgi \mgj \left \{ 2(\dotp13 \dotp25+\dotp15 \dotp23) - \dotp35 \, s 
\right \} \non \\ \non \\
K_{23}^a &=& 2 (\dotp13 \dotp25 + \dotp15 \dotp23) \dotp45 \non \\
K_{23}^b &=& \mgi \mgj  \left \{ s (\dotp15 + \dotp25)  - 4 \dotp15 \dotp25 \right \} \non \\
K_{23}^c &=& \left \{  s (\dotp14+\dotp24) +2 (\dotp14 \dotp25 + \dotp15 \dotp24) \right \}
\dotp35 - 4 \dotp15 \dotp25 \dotp34 \non \\
K_{23}^{c^\prime} &=& K_{23}^c - 2 \, s \, \dotp35 \dotp45 \non \\
K_{23}^d &=& 2 \dotp25 (\dotp13 \dotp45 - \dotp14 \dotp35) + {K_{23}^d}^\prime
+ {K_{23}^d}^{\prime\prime} \non \\
K_{23}^{d^\prime} &=& s \, \mgi \mgj (\dotp15-\dotp25) \non \\
K_{23}^{d^{\prime\prime}} &=& s (\dotp14-\dotp24) \dotp35 
\eeq

\nn
In the above expressions, $h$ is the suffix for the chirality ($L$ or $R$) of
the final state sfermion and $h^\prime$ is the other chirality ($R$ or $L$);
the latter suffix appears only on the couplings $O_{ij}$. $q=\p1+\p2$ is the 
sum of incoming 4-momenta. We used the notation $2\ve = C_L^e+C_R^e$ and $2\ae
= C_L^e-C_R^e$ and similarly for $\vf$ and $\af$. The $p_{ij}$ correspond to
the scalar products $p_i \cdot p_j$ and $s=q^2=(p_1+p_2)^2$. All other 
couplings and parameters are defined in the text. \s

Note that the antisymmetric part in each of the above terms, proportional to 
$a_e$ are pulled out apart and when, integrated over the full angular domain 
the corresponding contribution which antisymmetric, vanishes. \s

The differential cross section is obtained by dividing by the flux and 
multiplying by the phase space, 
\beq
{\rm d} \sigma = \frac{1}{2s} \times \frac{1}{(2 \pi)^5} \frac{ {\rm d}^3p_3}{ 
2E_3} \frac{ {\rm d}^3p_4}{ 2E_4} \frac{ {\rm d}^3p_5}{ 2E_5} \delta^4 (p_1+p_2
-p_3-p_4-p_5) \times |M|^2 
\eeq
The integral over the phase space, to obtain the total production cross 
section, is then performed numerically.


\newpage
 
\begin{thebibliography}{99} 
\bibitem{Tevatron} S. Abel et al., Report of the ``SUGRA" working group for 
``RUN II at the Tevatron", hep-ph/0003154; H. Baer, C.H. Chen, M. Drees, 
F. Paige and X. Tata, Phys. Rev. D58 (1998) 075008 and  D59 (1999) 055014.
%
\bibitem{LHC} CMS Collaboration (S. Abdullin et al.), CMS-NOTE-1998-006, 
hep--ph/9806366; D. Denegri, W. Majerotto and  L. Rurua, CMS-NOTE-1997-094, 
hep--ph/9711357; I. Hinchliffe et al., Phys. Rev. D55 (1997) 5520.
%
\bibitem{NLC} E. Accomando, Phys. Rept. 299 (1998) 1;  American Linear Collider
Working Group (T. Abe et al.),  Report SLAC-R-570 and hep-ex/0106057; J. Bagger
et al., hep-ex/0007022; H. Murayama and M. Peskin, Ann. Rev. Nucl. Part. Sci. 
46 (1996) 533. 
%
\bibitem{TESLA} TESLA Technical Design Report, Part I: ``Executive Summary",
F. Richard, J.R. Schneider, D. Trines and A. Wagner, TESLA report 2001--23; 
TESLA TDR, Part III: ``Physcis at $\ee$ Linear Collider"
D. Heuer, D. Miller, F. Richard and P.M. Zerwas (eds.) et al., Report 
DESY--01--011C, hep-ph/0106315.  
%
\bibitem{tests} For recent analyses, see e.g.: G.A. Blair, W. Porod and P.M. 
Zerwas, Phys.~Rev.~D63 (2001) 017703; J.L. Feng and M. Peskin, Phys.~Rev.~D64 
(2001) 115002; S.Y. Choi et al., Eur. Phys. J.~C8 (1999) 669 and Eur. Phys. J. 
C14 (2000) 535; S.Y. Choi et al., hep-ph/0108117; B. Allanach, D. Grellscheid 
and F. Quevedo hep-ph/0111057.  
%
\bibitem{MSSM} For reviews on the MSSM, see: P. Fayet and S. Ferrara,
Phys.  Rep. 32 (1977) 249; 
R. Barbieri, Riv. Nuov. Cim. 11 (1988) 1; R. Arnowitt and Pran Nath,
Report CTP-TAMU-52-93; M. Drees and S.P. Martin, in ``Electroweak
Symmetry Breaking and New Physics at the TeV Scale", eds. T. Barklow
et al., World Scientific (1996), hep-ph/9504324; S.P. Martin,
hep-ph/9709356; J. Bagger, Lectures at TASI-95, hep-ph/9604232;
H.P. Nilles, Phys. Rep. 110 (1984) 1.
%
\bibitem{HaberKane} H. E. Haber and G. Kane, Phys. Rep. 117 (1985) 75. 
%
\bibitem{mSUGRA} 
A.H. Chamseddine, R. Arnowitt and P. Nath, Phys. Rev. Lett. 49 (1982) 970;
R. Barbieri, S. Ferrara and C.A Savoy, Phys. Lett. B119 (1982) 343;
L. Hall, J. Lykken and S. Weinberg, Phys. Rev. D27 (1983) 2359. For recent
reviews see S. Abel et al. in Ref.~\cite{Tevatron} and A. Djouadi, M. Drees
and J.L. Kneur, JHEP 0108 (2001) 055.  
%
\bibitem{Asleptons}
A. Bartl, H. Fraas and W. Majerotto, Z. Phys. C34 (1987) 411; 
M. Chen, C. Dionisi, M. Martinez and X. Tata, Phys. Rep. 159 (1988) 201;
B. de Carlos amd M.A. Diaz, Phys. Lett. B417 (1998) 72.
%
\bibitem{Asquarks} K.I. Hikasa and M. Kobayashi, Phys. Rev. D36 (1987) 724; 
M. Drees and K. Hikasa, Phys. Lett. B252 (1990) 127;
%
\bibitem{Peter}
A. Freitas, D.J. Miller and P.M. Zerwas, Eur. Phys. J. C21 (2001) 361.
%
\bibitem{Manuel} M. Drees and O. Eboli, Eur. Phys. J. C10 (1999) 337. 
%
\bibitem{later} A. Datta, A. Djouadi and M. M\"uhlleitner, in preparation. 
%
\bibitem{HaberGunion}
J.F. Gunion and H.E. Haber, Nucl. Phys. B272 (1986) 1 and erratum 
hep-ph/9301205.
%
\bibitem{LEP2} For a summary on the experimental limits on the mass of
the Higgs and SUSY particles, see F. Gianotti, talk given at the 
EPS--HEP--2001, 12--18 July, Budapest. 
%
\bibitem{CompHep} A. Pukhov, E. Boos, M. Dubinin, V. Edneral, V. Ilyin, D.
Kovalenko, A. Kryukov, V. Savrin, S. Shichanin, and A. Semenov,  ``CompHEP - 
a package for evaluation of Feynman diagrams and integration over 
multi-particle phase space. User's manual for version 33", Preprint INP MSU 
98-41/542, hep-ph/9908288. 
%
\bibitem{gluino}
P. Nelson and P. Osland, Phys.~Lett.~B115 (1982) 407;
G.L. Kane and W.B. Rolnick, Nucl. Phys. B217 (1983) 117;
B.A. Campbell, J.A. Scott and M.K. Sundaresan, Phys. Lett. B126 (1993) 376;
B. Kileng and P. Osland, Z.~Phys.~C66 (1995) 503;
A. Djouadi and M. Drees,  Phys. Rev. D51 (1995) 4997.
%
\bibitem{Uli}  G. Blair and H.U. Martyn, hep--ph/9910416. 
%
 
\end{thebibliography}
\end{document}